\documentclass[aps,prd,superscriptaddress,showpacs,preprintnumbers]{revtex4}
\usepackage{graphicx,color}
\newcommand{\be}{\begin{equation}}
\newcommand{\ee}{\end{equation}}
\newcommand{\bea}{\begin{eqnarray}}
\newcommand{\eea}{\end{eqnarray}}

\newcommand{\bfk}{\mbox{\boldmath $k$}}

\def\kt{k_\perp}
\def\pt{p_\perp}
\newcommand{\bfp}{\mbox{\boldmath $p$}}

\newcommand{\bfq}{\mbox{\boldmath $q$}}
\newcommand{\bfP}{\mbox{\boldmath $P$}}
\newcommand{\bfS}{\mbox{\boldmath $S$}}

\newcommand{\pup}{p^\uparrow}

\newcommand{\qup}{q^\uparrow}

\newcommand{\pdown}{p^\downarrow}

\def\lsim{\mathrel{\rlap{\lower4pt\hbox{\hskip1pt$\sim$}}\raise1pt\hbox{$<$}}}
\def\gsim{\mathrel{\rlap{\lower4pt\hbox{\hskip1pt$\sim$}}\raise1pt\hbox{$>$}}}
\def\nostrocostruttino#1\over#2{\mathrel{\mathop{\kern 0pt \rlap
{\hbox{$#1$}}} \hbox{\kern-.135em $#2$}}}

\begin{document}

\title{Transverse single-spin asymmetries in proton-proton collisions at the\\
AFTER@LHC experiment in a TMD factorisation scheme}

\author{M.~Anselmino}
\affiliation{Dipartimento di Fisica, Universit\`a di Torino,
             Via P. Giuria 1, I-10125 Torino, Italy}
\affiliation{INFN, Sezione di Torino, Via P. Giuria 1, I-10125 Torino, Italy}
\author{U.~D'Alesio}
\affiliation{Dipartimento di Fisica, Universit\`a di Cagliari, Cittadella
             Universitaria, I-09042 Monserrato (CA), Italy}
\affiliation{INFN, Sezione di Cagliari,
             C.P. 170, I-09042 Monserrato (CA), Italy}
\author{S.~Melis}
\affiliation{Dipartimento di Fisica, Universit\`a di Torino,
             Via P. Giuria 1, I-10125 Torino, Italy}
\date{\today}

\begin{abstract}
The inclusive large-$p_T$ production of a single pion, jet or direct photon, and Drell-Yan processes, are considered for proton-proton collisions in the kinematical range expected for the fixed-target experiment AFTER,
proposed at LHC. For all these processes, predictions are given for the transverse
single-spin asymmetry, $A_N$, computed according to a Generalised Parton Model
previously discussed in the literature and based on TMD factorisation. Comparisons
with the results of a collinear twist-3 approach, recently presented, are made and
discussed.
\end{abstract}

\pacs{13.88.+e, 13.85.Ni, 13.85.Qk}

\maketitle

\section{\label{1}Introduction and formalism}

Transverse Single-Spin Asymmetries (TSSAs), have been abundantly observed
in several inclusive proton-proton experiments since a long time; when
reaching large enough energies and $p_T$ values, their understanding from
basic quark-gluon QCD interactions is a difficult and fascinating task,
which has always been one of the major challenges for QCD.

In fact, large TSSAs cannot be generated by the hard elementary processes,
because of helicity conservation (in the massless limit) typical of QED and
QCD interactions; indeed, such asymmetries were expected to vanish at high
energies. Their persisting must be related to non perturbative properties of
the nucleon structure, such as parton intrinsic and orbital motion. A true
understanding of the origin of TSSAs would allow a deeper understanding of the
nucleon structure.

Since the 1990s two different, although somewhat related, approaches have attempted
to tackle the problem. One is based on the collinear QCD factorisation scheme and
involves as basic quantities, which can generate single spin dependences, higher-
twist quark-gluon-quark correlations in the nucleon as well as higher-twist fragmentation correlators.
The second approach is based on a physical, although unproven, generalisation of the
parton model, with the inclusion, in the factorisation scheme, of transverse momentum
dependent partonic distribution and fragmentation functions (TMDs), which also can
generate single spin dependences. The twist-3 correlations are related to moments of
some TMDs. We refer to Refs.~\cite{Anselmino:2012rq, Anselmino:2013rya,
Kanazawa:2014nea, Kanazawa:2014dca, Qiu:1991wg, Qiu:1998ia, Kouvaris:2006zy,
Ji:2006vf, D'Alesio:2007jt}, and references therein, for a more detailed account
of the two approaches, and possible variations, with all relevant citations.
Following Ref.~\cite{Kanazawa:2015fia}, we denote by CT-3 the first approach
while the second one is, as usual, denoted by GPM.

In this paper we consider TSSAs at the proposed AFTER@LHC experiment, in which
high-energy protons extracted from the LHC beam would collide on a (polarised)
fixed target of protons, with high luminosity. For a description of the physics
potentiality of this experiment see Ref.~\cite{Brodsky:2012vg} and for the latest
technical details and importance for TMD studies see, for example,
Ref.~\cite{Massacrier:2015nsm}. Due to its features the AFTER@LHC is an ideal
experiment to study and understand the origin of SSAs and, in general,
the role of QCD interactions in high-energy hadronic collisions;
AFTER@LHC would be a polarised fixed target experiment with unprecedented high
luminosity.

We recall our formalism by considering the Transverse Single-Spin Asymmetry $A_N$, measured in $p \, \pup \to h \, X$ inclusive reactions and defined as:
\be
A_N = \frac{d\sigma^\uparrow - d\sigma^\downarrow}
           {d\sigma^\uparrow + d\sigma^\downarrow}
\quad\quad {\rm with} \quad\quad
d\sigma^{\uparrow, \downarrow} \equiv
\frac{E_h \, d\sigma^{p \, p^{\uparrow, \downarrow} \to h \, X}}
{d^{3} \bfp_h} \>, \label{an}
\ee
where $\uparrow, \downarrow$ are opposite spin orientations perpendicular to the
$x$-$z$ scattering plane, in the $p \, \pup$ c.m. frame. We define the $\uparrow$
direction as the $+\hat y$-axis and the unpolarised proton is moving
along the $+\hat z$-direction. In such a process the only large scale is the
transverse momentum $p_T=|(\bfp_h)_x|$ of the final hadron.

In the GPM $A_N$ originates mainly from two spin and transverse momentum effects,
one introduced by Sivers in the partonic distributions~\cite{Sivers:1989cc,
Sivers:1990fh}, and one by Collins in the parton fragmentation
process~\cite{Collins:1992kk}, being all the other effects strongly
suppressed by azimuthal phase integrations~\cite{Anselmino:2005sh}.
According to the Sivers effect the number density of unpolarised quarks $q$ (or
gluons) with intrinsic transverse momentum $\bfk_\perp$ inside a transversely
polarised proton $\pup$, with
three-momentum $\bfP$ and spin polarisation vector $\bfS$, can be written as
\be
\hat f_ {q/\pup} (x,\bfk_\perp) = f_ {q/p} (x,\kt) +
\frac{1}{2} \, \Delta^N \! f_ {q/\pup}(x,\kt) \;
{\bfS} \cdot (\hat {\bfP}  \times \hat{\bfk}_\perp)
\,,\label{sivnoi}
\ee
where $x$ is the proton light-cone momentum fraction carried by the quark,
$f_ {q/p}(x,\kt)$ is the unpolarised TMD ($\kt = |\bfk_\perp|$) and
$\Delta^N \! f_ {q/\pup}(x,\kt)$ is the Sivers function. $\hat {\bfP} =
\bfP/|\bfP|$ and $\hat{\bfk}_\perp = \bfk_\perp/\kt$ are unit vectors.
Notice that the Sivers function is most often denoted as
$f_{1T}^{\perp q}(x, k_\perp)$~\cite{Mulders:1995dh}; this notation is
related to ours by~\cite{Bacchetta:2004jz}
\be
\Delta^N \! f_ {q/\pup}(x,k_\perp) = - \frac{2\,k_\perp}{m_p} \>
f_{1T}^{\perp q}(x, k_\perp) \>, \label{rel}
\ee
where $m_p$ is the proton mass.

Similarly, according to the Collins effect the number density of unpolarised
hadrons $h$ with transverse momentum $\bfp_\perp$ resulting in the fragmentation
of a transversely polarised quark $\qup$, with three-momentum $\bfq$ and spin polarisation vector $\bfS_q$, can be written as
\be
\hat D_{\qup/h} (z,\bfp_\perp) = D_{h/q} (z,\pt) +
\frac{1}{2} \, \Delta^N \! D_{\qup/h}(z,\pt) \;
{\bfS_q} \cdot (\hat {\bfq}  \times \hat{\bfp}_\perp)
\,,\label{colnoi}
\ee
where $z$ is the parton light-cone momentum fraction carried by the hadron,
$D_ {h/q}(z,\pt)$ is the unpolarised TMD ($\pt = |\bfp_\perp|$) and
$\Delta^N \! D_ {\qup/h}(z,\pt)$ is the Collins function. $\hat {\bfq} =
\bfq/|\bfq|$ and $\hat{\bfp}_\perp = \bfp_\perp/\pt$ are unit vectors.
Notice that the Collins function is most often denoted as
$H_{1}^{\perp q}(z, p_\perp)$~\cite{Mulders:1995dh}; this notation is
related to ours by~\cite{Bacchetta:2004jz}
\be
\Delta^N D_{h/\qup}(z, \pt) = \frac{2\,p_\perp}{zM_h} \>
H_{1}^{\perp q}(z,\pt) \>, \label{rel1}
\ee
where $M_h$ is the hadron mass.

According to the GPM formalism~\cite{Anselmino:2005sh, Anselmino:2012rq,
Anselmino:2013rya}, $A_N$ can then be written as:
\be
A_N = \frac{[d\sigma^\uparrow - d\sigma^\downarrow]_{\rm Sivers}
+ [d\sigma^\uparrow - d\sigma^\downarrow]_{\rm Collins}}
{d\sigma^\uparrow + d\sigma^\downarrow} \>\cdot \label{ansc}
\ee
The Collins and Sivers contributions were recently studied, respectively in
Refs.~\cite{Anselmino:2012rq} and \cite{Anselmino:2013rya}, and are given by:
\bea
[d\sigma^\uparrow - d\sigma^\downarrow]_{\rm Sivers}
&=& \!\!\! \sum_{a,b,c,d} \int \frac{dx_a \, dx_b \, dz}
{16 \, \pi^2 \, x_a \, x_b \, z^2 s} \; d^2 \bfk_{\perp a} \,
d^2 \bfk_{\perp b}\, d^3 \bfp_{\perp}\,
\delta(\bfp_\perp \cdot \hat{\bfp}_c) \> J(p_{\perp}) \>
\delta(\hat s + \hat t + \hat u) \nonumber\\
&\times& \Delta^N\!f_{a/\pup}(x_a, k_{\perp a}) \,
\cos (\phi_a) \, f_{b/p}(x_b, k_{\perp b}) \> \frac{1}{2}
\left[ |\hat M_1^0|^2 + |\hat M_2^0|^2 + |\hat M_3^0|^2 \right]_{ab\to cd} \>
D_{h/c}(z, p_\perp) \>, \label{numans}
\eea
and
\bea
[d\sigma^\uparrow - d\sigma^\downarrow]_{\rm Collins}
&=& \!\!\! \sum_{q_a,b,q_c,d} \int \frac{dx_a \, dx_b \, dz}
{16 \, \pi^2 \, x_a \, x_b \, z^2 s} \; d^2 \bfk_{\perp a} \,
d^2 \bfk_{\perp b}\, d^3 \bfp_{\perp}\,
\delta(\bfp_\perp \cdot \hat{\bfp}_c) \> J(p_{\perp}) \>
\delta(\hat s + \hat t + \hat u) \nonumber \\
&\times& \Delta_Tq_a(x_a, k_{\perp a}) \,
\cos (\phi_a + \varphi_1 - \varphi_ 2 + \phi_\pi^H) \label{numanc} \\
&\times& f_{b/p}(x_b, k_{\perp b}) \>
\left[ \hat M_1^0 \, \hat M_2^0 \right]_{q_ab\to q_cd} \>
\Delta^N D_{h/\qup_c}(z, p_\perp) \>. \nonumber
\eea

For details and a full explanation of the notations in the above equations
we refer to Ref.~\cite{Anselmino:2005sh} (where $\bfp_\perp$ is denoted as
$\bfk_{\perp C}$). It suffices to notice here that $J(p_{\perp})$ is a
kinematical factor, which at ${\cal O}(p_\perp/E_{h})$ equals 1. The phase
factor $\cos(\phi_a)$ in Eq.~(\ref{numans}) originates directly from the
$\bfk_\perp$ dependence of the Sivers distribution [${\bfS} \cdot (\hat {\bfP}
\times \hat{\bfk}_\perp)$, Eq.~(\ref{sivnoi})]. The (suppressing) phase factor
$\cos(\phi_a + \varphi_1 - \varphi_ 2 + \phi_\pi^H)$ in Eq.~(\ref{numanc})
originates from the $\bfk_\perp$ dependence of the unintegrated transversity
distribution $\Delta_Tq$, the polarized elementary interaction and the
spin-$\bfp_\perp$ correlation in the Collins function. The explicit expressions
of $\varphi_1, \varphi_2$ and $\phi_\pi^H$ in terms of the integration variables
can be found via Eqs.~(60)-(63) in~\cite{Anselmino:2005sh} and Eqs.~(35)-(42)
in~\cite{Anselmino:2004ky}.

The $\hat M_i^0$'s are the three independent hard scattering helicity amplitudes
describing the lowest order QCD interactions. The sum of their moduli squared
is related to the elementary unpolarised cross section
$d\hat\sigma^{a b \to c d}$, that is
\be
\label{eq:sigma}
\frac{d\hat\sigma^{ab\to cd}}{d\hat t} = \frac{1}{16\pi\hat s^2}\,\frac{1}{2} \sum_{i=1}^3 |\hat M_i^0|^2\,.
\ee
The explicit expressions of the combinations of $\hat M_i^0$'s which give the
QCD dynamics in Eqs.~(\ref{numans}) and (\ref{numanc}), can be found, for all
possible elementary interactions, in Ref.~\cite{Anselmino:2005sh} (see also
Ref.~\cite{Anselmino:2012rq} for a correction to one of the product of
amplitudes). The QCD scale is chosen as $Q = p_T$.

The denominator of Eq.~(\ref{an}) or ~(\ref{ansc}) is twice the unpolarised
cross section and is given in our TMD factorisation by the same expression
as in Eq.~(\ref{numans}), where one simply replaces the factor
$\Delta^N\!f_{a/\pup}\,\cos(\phi_a)$ with $2f_{a/p}$.

\section{$A_N$ for single pion, jet and direct photon production}

We present here our results for $A_N$, Eq.~(\ref{an}), based on our GPM scheme,
Eqs.~(\ref{ansc}), (\ref{numans}) and (\ref{numanc}). The TMDs which enter in
these equations are those extracted from the analysis of Semi Inclusive Deep
Inelastic (SIDIS) and $e^+e^-$ data~\cite{Anselmino:2005ea, Anselmino:2008sga,
Anselmino:2007fs, Anselmino:2008jk}, adopting simple factorised forms, which
we recall here. For the unpolarised TMD partonic distributions and fragmentation
functions we have, respectively:
\be
f_{q/p}(x,k_\perp) = f_{q/p}(x)\,
\frac{e^{-k_\perp^2/\langle k_\perp^2 \rangle}}
{\pi \langle k_\perp^2 \rangle}
\quad\quad\quad \langle k_\perp^2\rangle = 0.25\, {\rm GeV}^2
\label{TMDpdf}
\ee
and
\be
D_{h/q}(z,p_\perp) = D_{h/q}(z)\,
\frac{e^{-p_\perp^2/\langle p_\perp^2 \rangle}}
{\pi \langle p_\perp^2 \rangle}
\quad\quad\quad \langle p_\perp^2\rangle = 0.20\, {\rm GeV}^2 \>.
\label{TMDff}
\ee
The Sivers function is parameterised as
\be
\Delta^N\! f_{q/p^\uparrow}(x,k_\perp) = 2 \, {\cal N}_q^S(x)\,f_{q/p}(x)\,
h(k_\perp)\,\frac{e^{-k_\perp^2/\langle k_\perp^2 \rangle}}
{\pi \langle k_\perp^2 \rangle}\,,
\label{eq:siv-par}
\ee
where
\be
{\cal N}_q^S(x) = N_q^S x^{\alpha_q}(1-x)^{\beta_q}\,
\frac{(\alpha_q+\beta_q)^{(\alpha_q+\beta_q)}}
{\alpha_q^{\alpha_q}\beta_q^{\beta_q}}\,,
\label{eq:nq-siv}
\ee
with $|N_q^{S}|\leq 1$, and
\be
h(k_\perp) = \sqrt{2e}\,\frac{k_\perp}{M}\,e^{-k_\perp^2/M^2}\,.
\label{eq:h-siv}
\ee
Similarly, the quark transversity distribution, $\Delta_T q(x,k_\perp)$, and
the Collins fragmentation function, $\Delta^N D_{h/q^\uparrow}(z,p_\perp)$,
have been parametrized as follows:
\be
\Delta_T q(x,k_\perp) = \frac{1}{2}\,{\cal N}_q^T(x)\,\left[\,f_{q/p}(x)+
\Delta q(x)\,\right]\,\frac{e^{-k_\perp^2/\langle k_\perp^2 \rangle}}
{\pi \langle k_\perp^2 \rangle}\,,
\label{eq:transv-par}
\ee
\be
\Delta^N\! D_{h/q^\uparrow}(z,p_\perp) = 2 {\cal N}_q^C(z)\,D_{h/q}(z)\,
h(p_\perp)\,\frac{e^{-p_\perp^2/\langle p_\perp^2 \rangle}}
{\pi \langle p_\perp^2 \rangle}\,,
\label{eq:coll-par}
\ee
where $\Delta q(x)$ is the usual collinear quark helicity distribution,
\be
{\cal N}_q^T(x) = N_q^T x^{a_q}(1-x)^{b_q}\,
\frac{(a_q + b_q)^{(a_q + b_q)}}
{a_q^{a_q}a_q^{b_q}}\,,
\label{eq:nq-trans}
\ee
\be
{\cal N}_q^C(z) = N_q^C z^{\gamma_q}(1-z)^{\delta_q}\,
\frac{(\gamma_q+\delta_q)^{(\gamma_q+\delta_q)}}
{\gamma_q^{\gamma_q}\delta_q^{\delta_q}}\,,
\label{eq:nq-coll}
\end{equation}
with $|N_q^{T(C)}|\leq 1$, and
\begin{equation}
h(p_\perp) = \sqrt{2e}\,\frac{p_\perp}{M_c}\,e^{-p_\perp^2/M_c^2}\,.
\label{eq:h-coll}
\ee

All details concerning the motivations for such a choice, the values of the
parameters and their derivation can be found in Refs.~\cite{Anselmino:2005ea,
Anselmino:2008sga, Anselmino:2007fs, Anselmino:2008jk}. We do not repeat
them here, but in the caption of each figure we will give the corresponding
references which allow to fix all necessary values.
\begin{figure}[ht!]
\includegraphics[width=8.5truecm,angle=0]{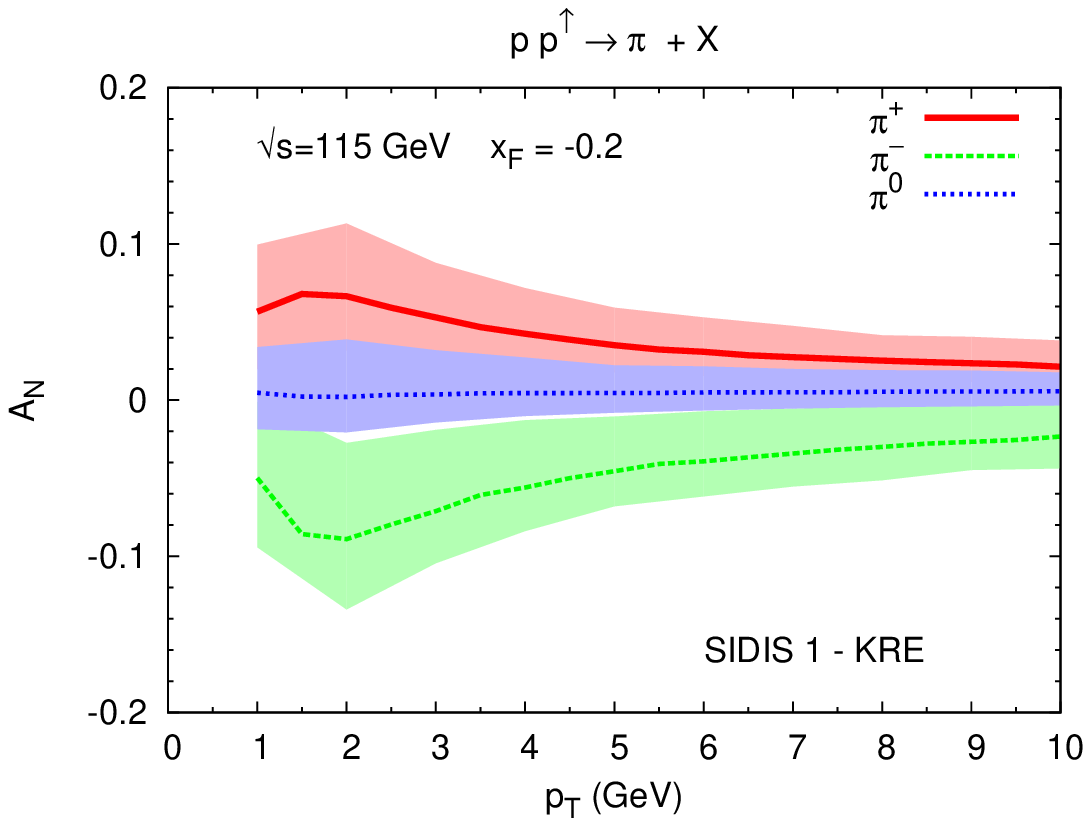}
\includegraphics[width=8.5truecm,angle=0]{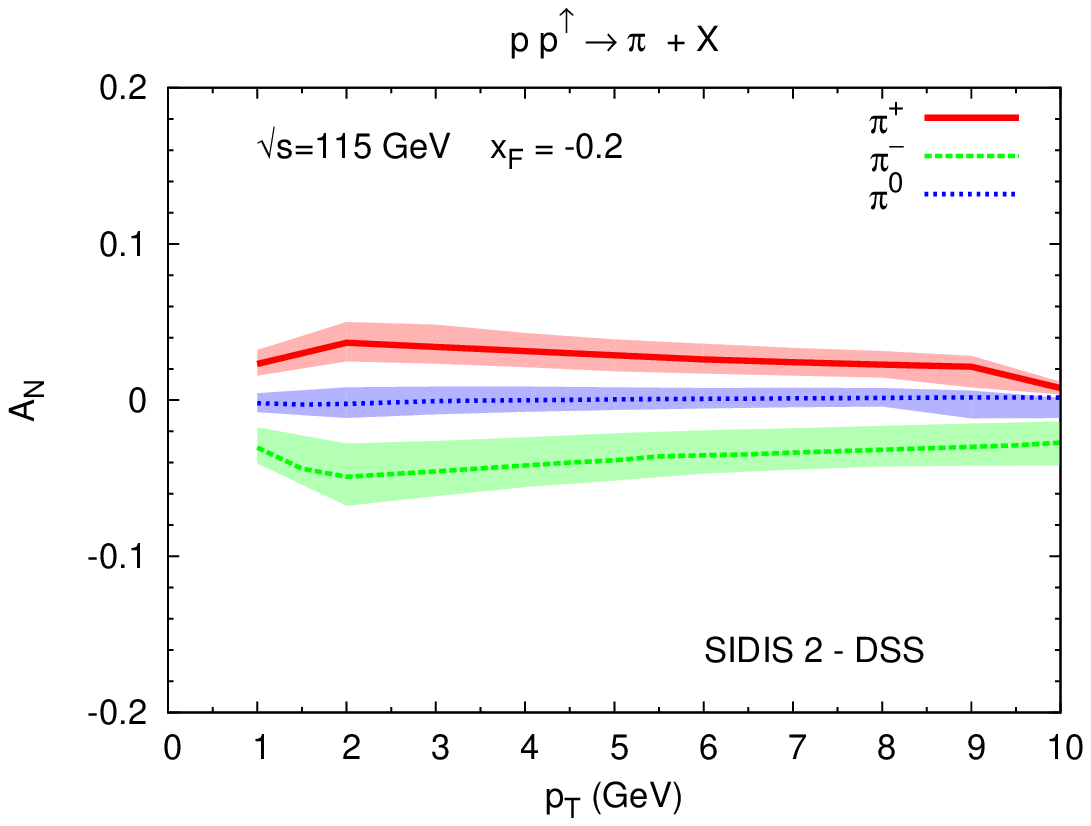}
\vskip 0.15 truecm
\includegraphics[width=8.5truecm,angle=0]{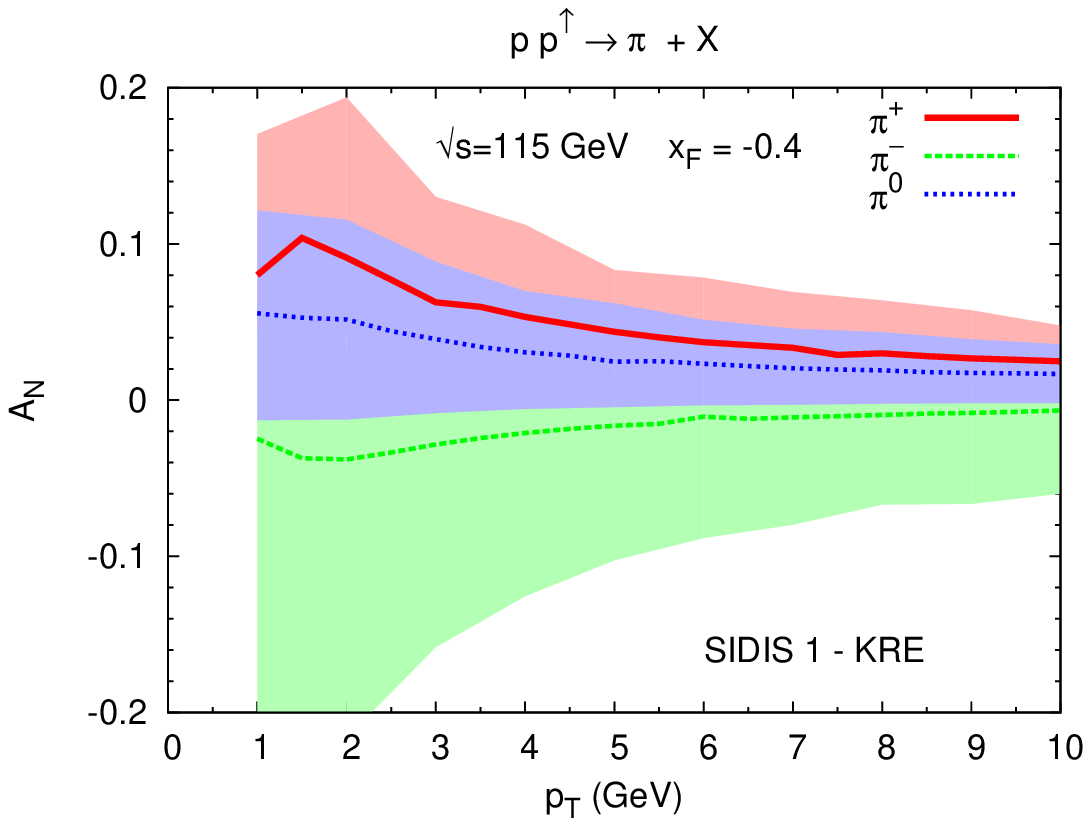}
\includegraphics[width=8.5truecm,angle=0]{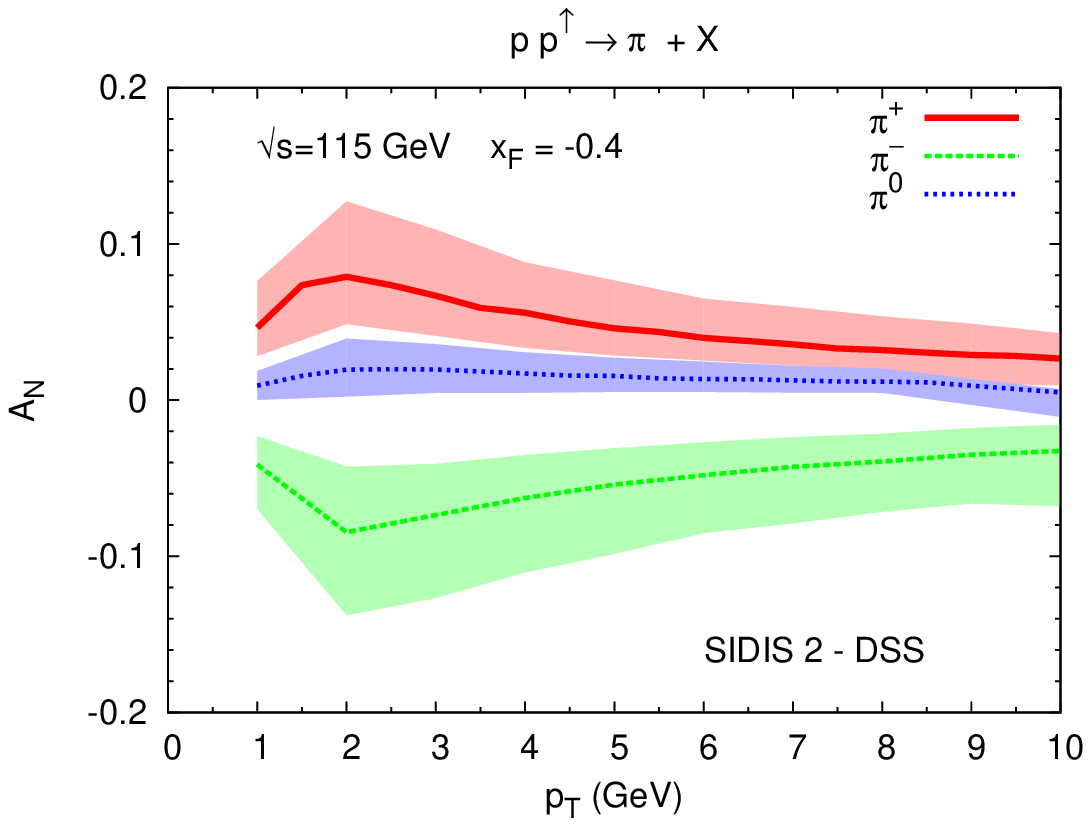}
\caption{Our theoretical estimates for $A_N$ vs.~$p_T$ at $\sqrt{s} = 115$ GeV,
$x_F = -0.2$ (upper plots) and $x_F = -0.4$ (lower plots) for inclusive
$\pi^\pm$ and $\pi^0$ production in $p \, \pup \to \pi \, X$ processes,
computed according to Eqs~(\ref{ansc})--(\ref{numanc}) of the text.
The contributions from the Sivers and the Collins effects are added together.
The computation is performed adopting the Sivers and Collins functions of
Refs.~\cite{Anselmino:2005ea, Anselmino:2007fs} (SIDIS 1 - KRE, left panels),
and of Refs.~\cite{Anselmino:2008sga, Anselmino:2008jk} (SIDIS 2 - DSS, right
panels). The overall statistical uncertainty band, also shown, is the envelope
of the two independent statistical uncertainty bands obtained following the
procedure described in Appendix A of Ref.~\cite{Anselmino:2008sga}.}
\label{fig1}
\end{figure}

We present our results on $A_N$ for the process $p \, \pup \to \pi \, X$
at the expected AFTER@LHC energy ($\sqrt s = 115$ GeV) in
Figs.~\ref{fig1}-\ref{fig3}. Following Refs.~\cite{Anselmino:2012rq,
Anselmino:2013rya}, our results are given for two possible choices of the
SIDIS TMDs, and are shown as function of $p_T$ at two fixed $x_F$ values
(Fig.~\ref{fig1}), as function of $x_F$ at two fixed rapidity $y$ values
(Fig.~\ref{fig2}) and as function of rapidity at one fixed $p_T$ value
(Fig.~\ref{fig3}). $x_F$ is the usual Feynman variable defined as
$x_F = 2p_L/{\sqrt s}$ where $p_L = (\bfp_h)_z$ is the $z$-component of
the final hadron momentum. Notice that, in our chosen reference frame, a
forward production, with respect to the polarised proton, means negative
values of $x_F$. The uncertainty bands reflects the uncertainty in the
determinations of the TMDs and are computed according to the procedure
explained in the Appendix of Ref.~\cite{Anselmino:2008sga}. More information
can be found in the figure captions.

Notice that, for both our choices of the Sivers functions,
the gluon Sivers distributions are taken to be vanishing, as suggested by
data~\cite{Anselmino:2008sga,Anselmino:2006yq}.
Gluon channels contribute instead to the unpolarised cross sections, in the
denominator of Eq.~(\ref{an}) or ~(\ref{ansc}). For the unpolarised partonic
distributions we adopt the GRV98LO PDF set~\cite{Gluck:1998xa} and for the
fragmentation functions the DSS set from Ref.~\cite{deFlorian:2007aj} and the Kretzer (KRE) set from Ref.~\cite{Kretzer:2000yf}.

\begin{figure}[h!t]
\includegraphics[width=8.5truecm,angle=0]{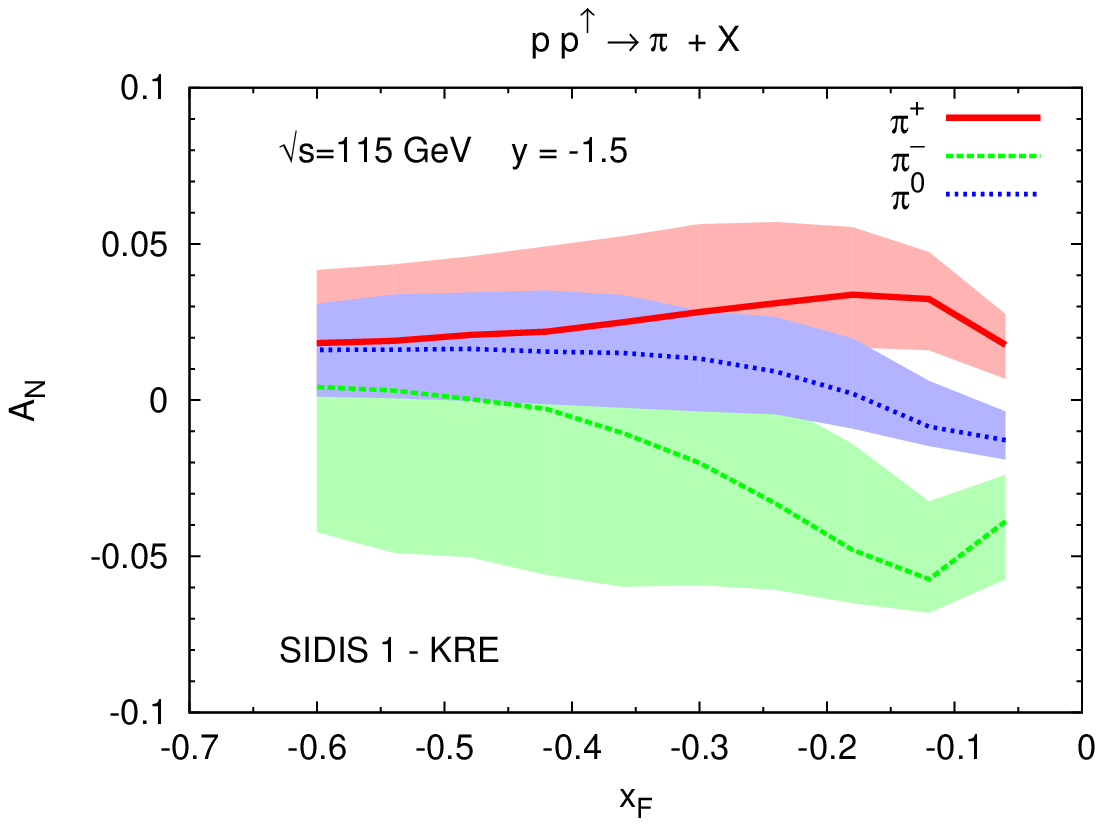}
\includegraphics[width=8.5truecm,angle=0]{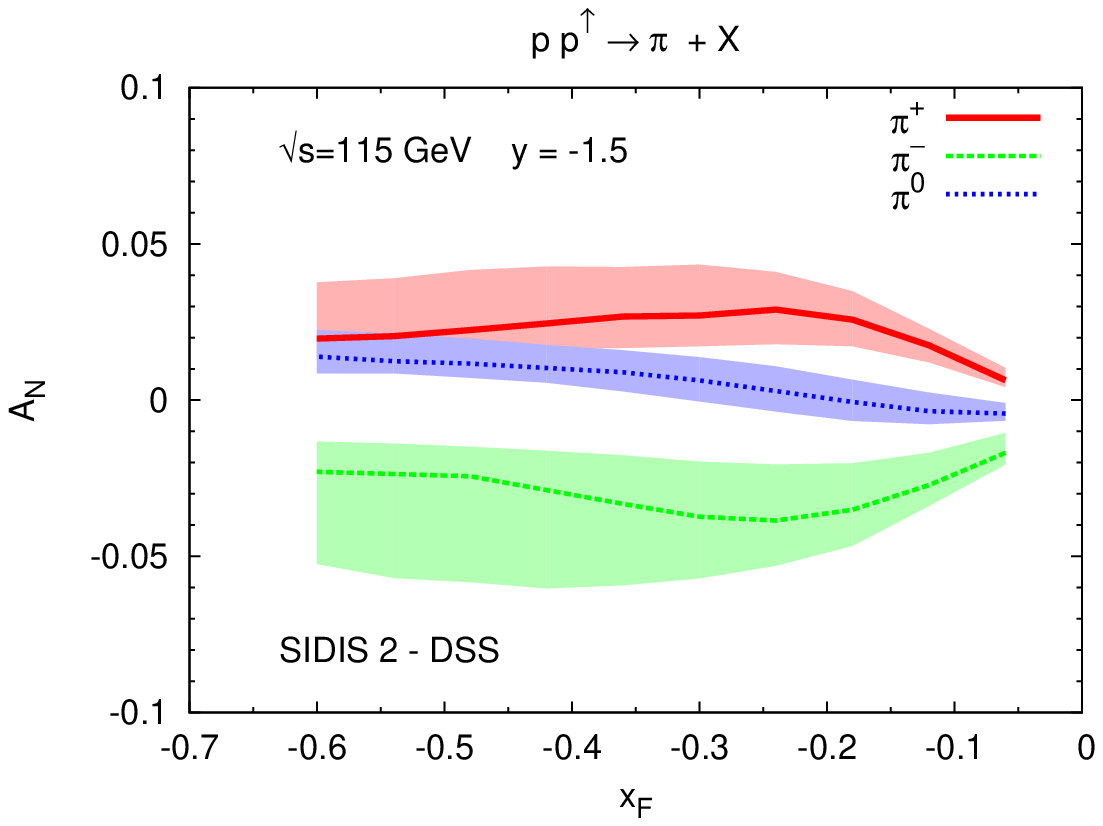}
\vskip 0.15 truecm
\includegraphics[width=8.5truecm,angle=0]{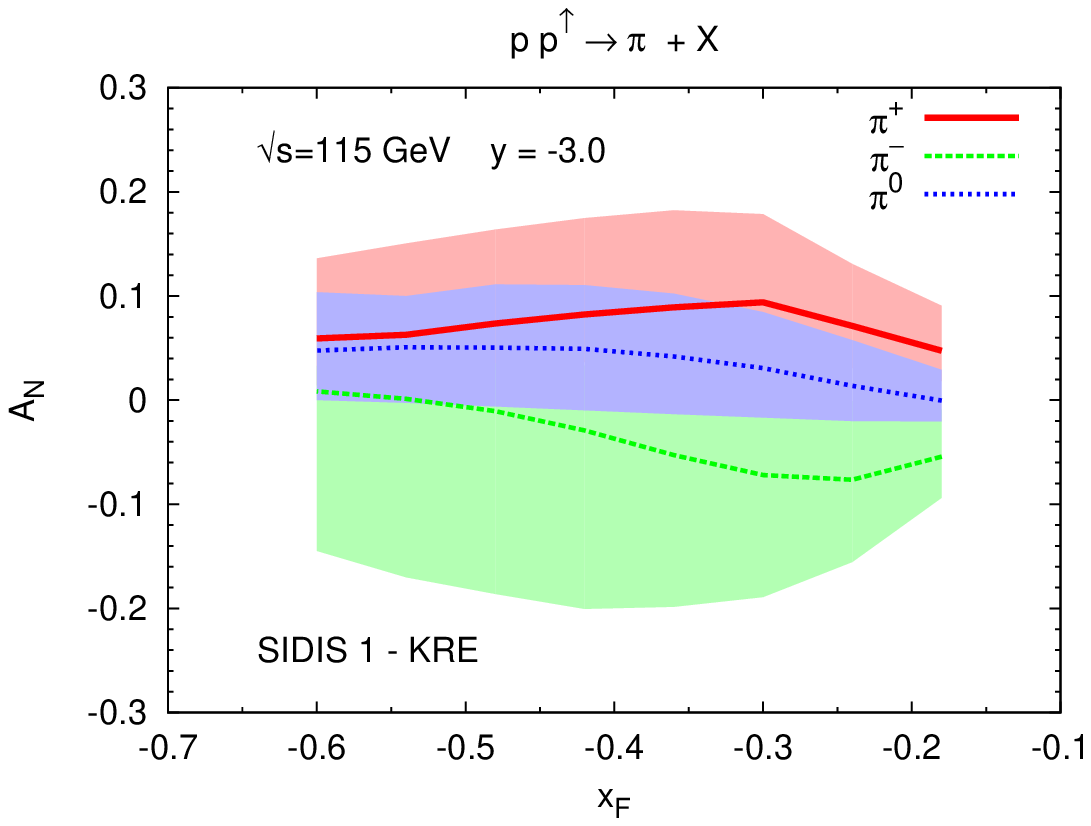}
\includegraphics[width=8.5truecm,angle=0]{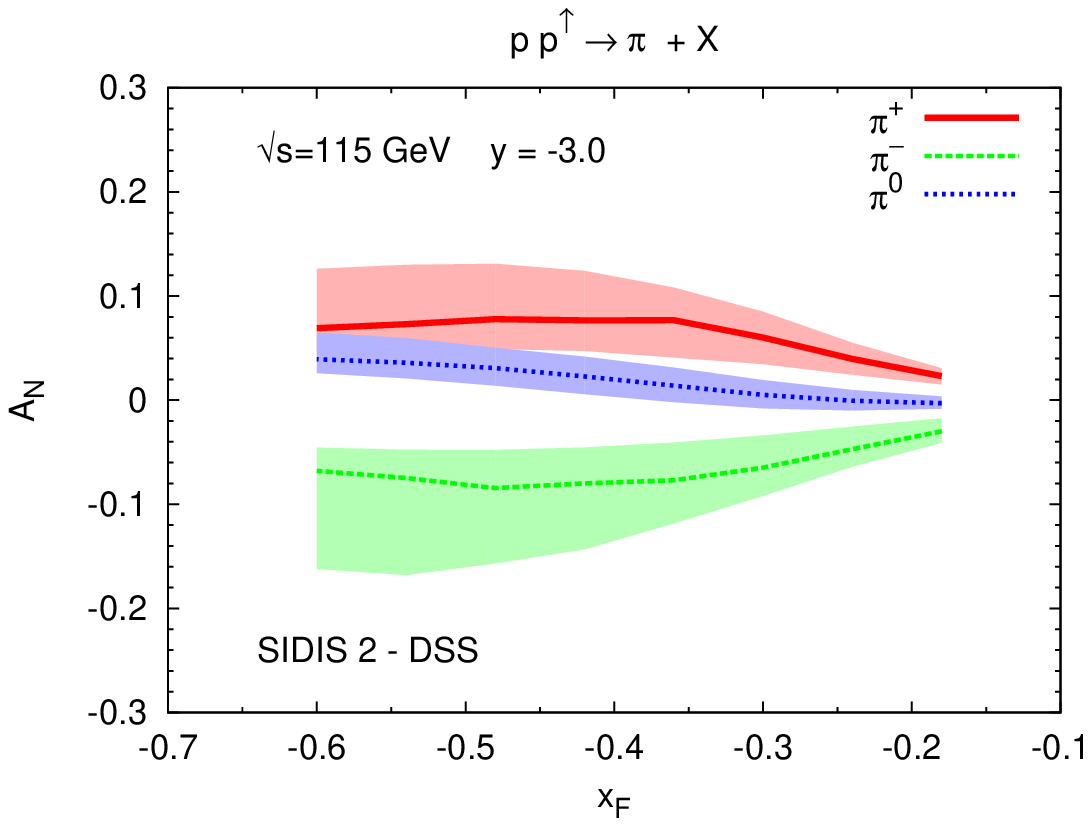}
\caption{Our theoretical estimates for $A_N$ vs.~$x_F$ at $\sqrt{s} = 115$ GeV,
$y = -1.5$ (upper plots) and $y = -3.0$ (lower plots) for inclusive
$\pi^\pm$ and $\pi^0$ production in $p \, \pup \to \pi \, X$ processes, computed
according to Eqs~(\ref{ansc})--(\ref{numanc}) of the text.
The contributions from the Sivers and the Collins effects are added together.
The computation is performed adopting the Sivers and Collins functions of
Refs.~\cite{Anselmino:2005ea, Anselmino:2007fs} (SIDIS 1 - KRE, left panels),
and of Refs.~\cite{Anselmino:2008sga, Anselmino:2008jk} (SIDIS 2 - DSS, right
panels). The overall statistical uncertainty band, also shown, is the envelope
of the two independent statistical uncertainty bands obtained following the
procedure described in Appendix A of Ref.~\cite{Anselmino:2008sga}.}
\label{fig2}
\end{figure}
\begin{figure}[h!t]
\includegraphics[width=8.5truecm,angle=0]{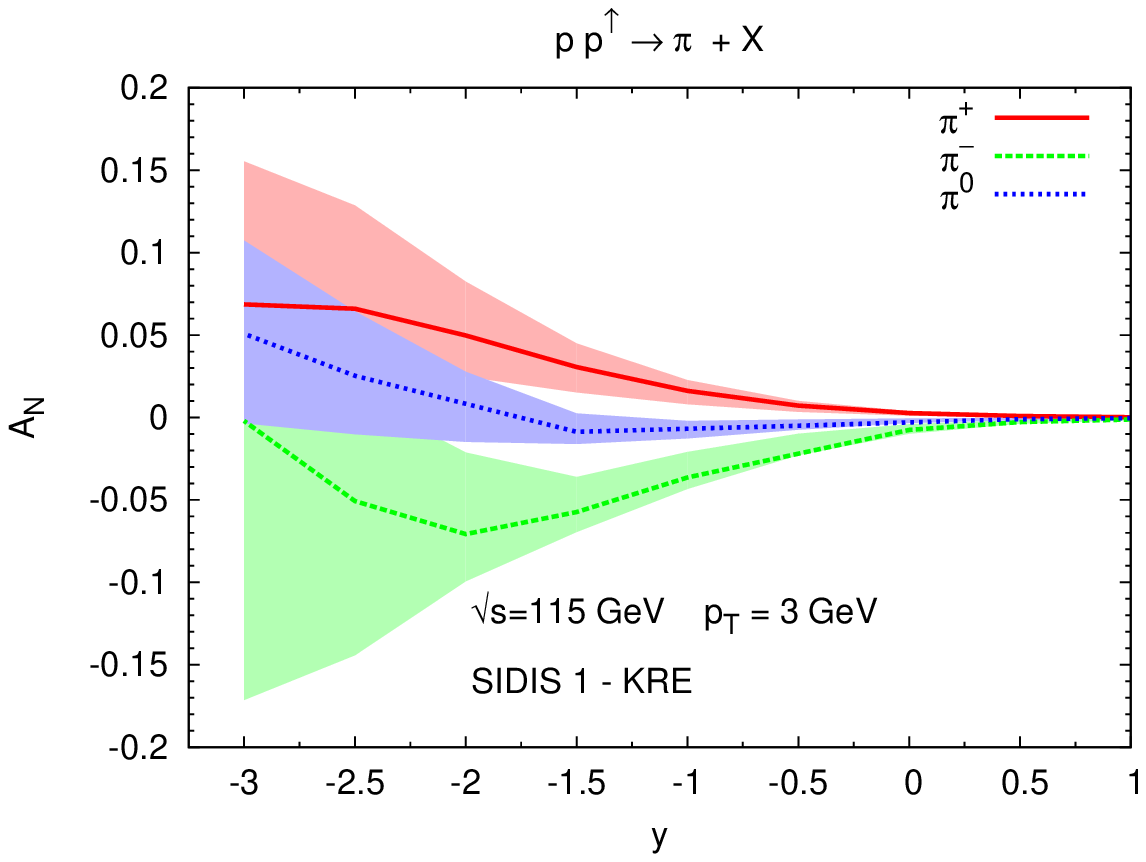}
\includegraphics[width=8.5truecm,angle=0]{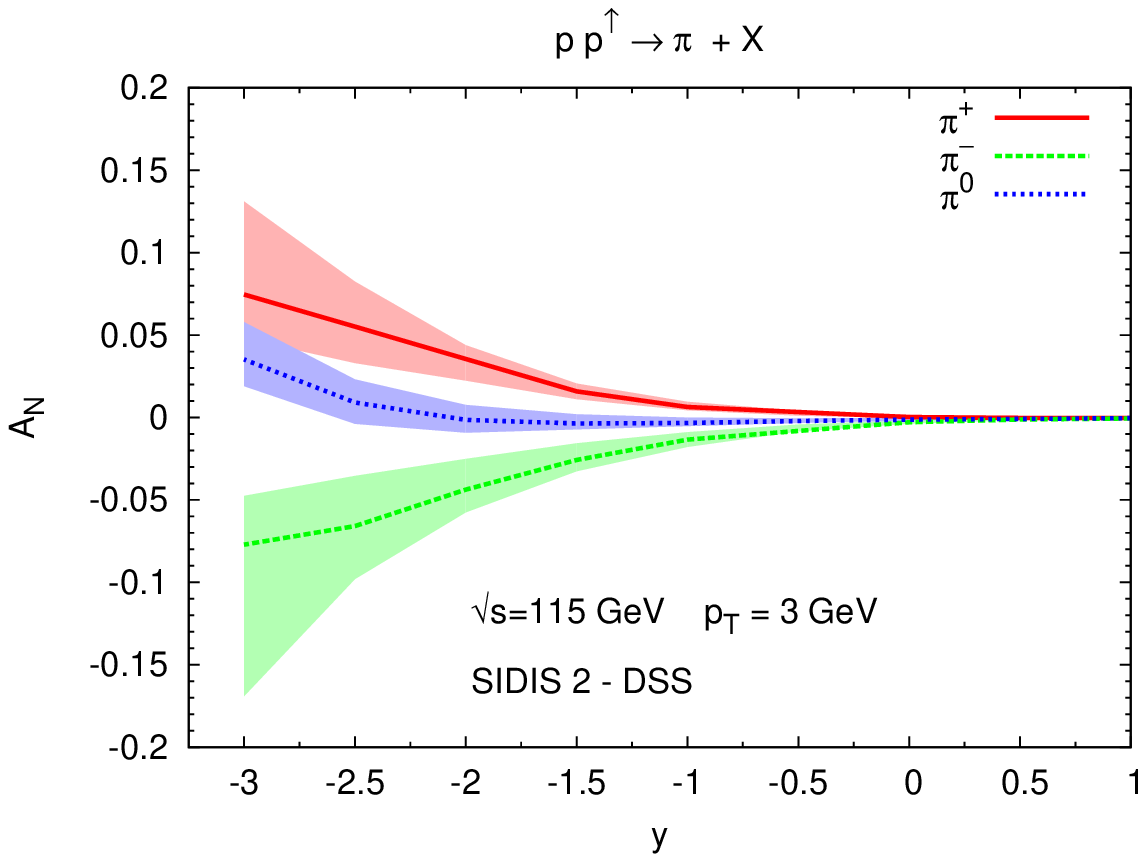}
\caption{Our theoretical estimates for $A_N$ vs.~$y$ at $\sqrt{s} = 115$ GeV
and $p_T = 3$ GeV, for inclusive $\pi^\pm$ and $\pi^0$ production in
$p \, \pup \to \pi \, X$ processes, computed according to
Eqs~(\ref{ansc})--(\ref{numanc}) of the text.
The contributions from the Sivers and the Collins effects are added together.
The computation is performed adopting the Sivers and Collins functions of
Refs.~\cite{Anselmino:2005ea, Anselmino:2007fs} (SIDIS 1 - KRE, left panel),
and of Refs.~\cite{Anselmino:2008sga, Anselmino:2008jk} (SIDIS 2 - DSS, right
panel). The overall statistical uncertainty band, also shown, is the envelope
of the two independent statistical uncertainty bands obtained following the
procedure described in Appendix A of Ref.~\cite{Anselmino:2008sga}.}
\label{fig3}
\end{figure}
\begin{figure}[h!t]
\includegraphics[width=8.5truecm,angle=0]{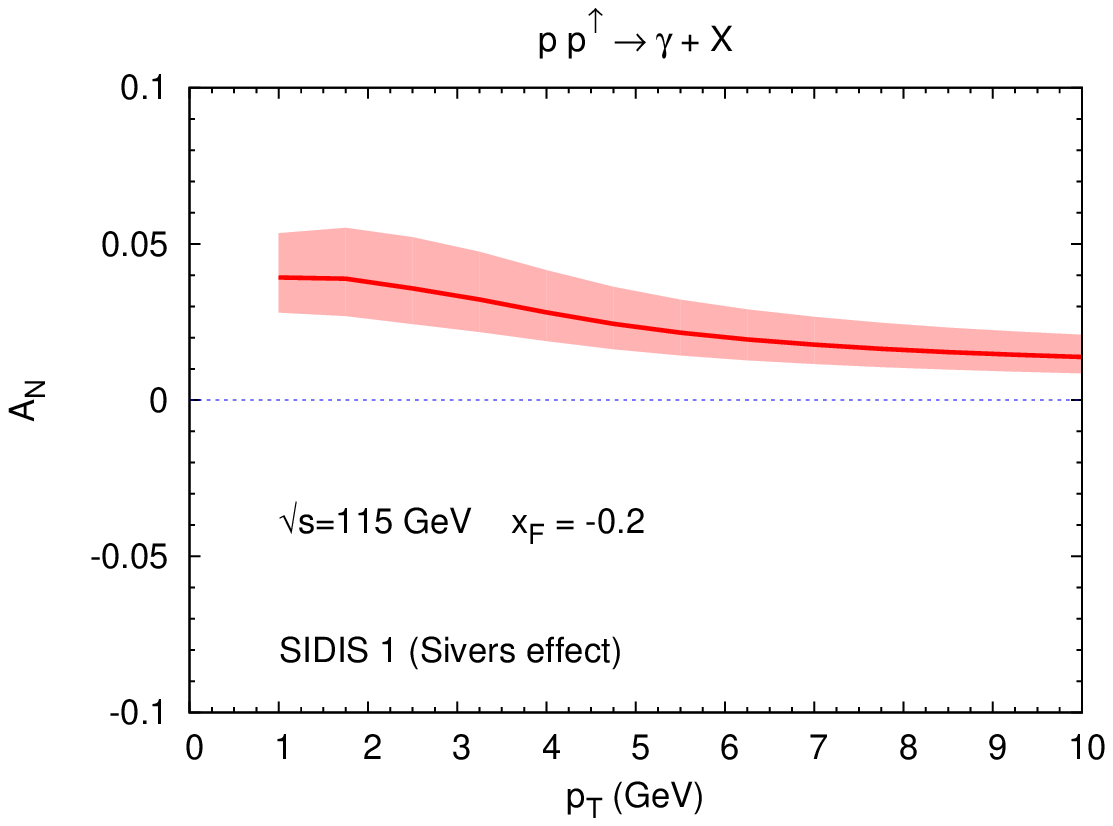}
\includegraphics[width=8.5truecm,angle=0]{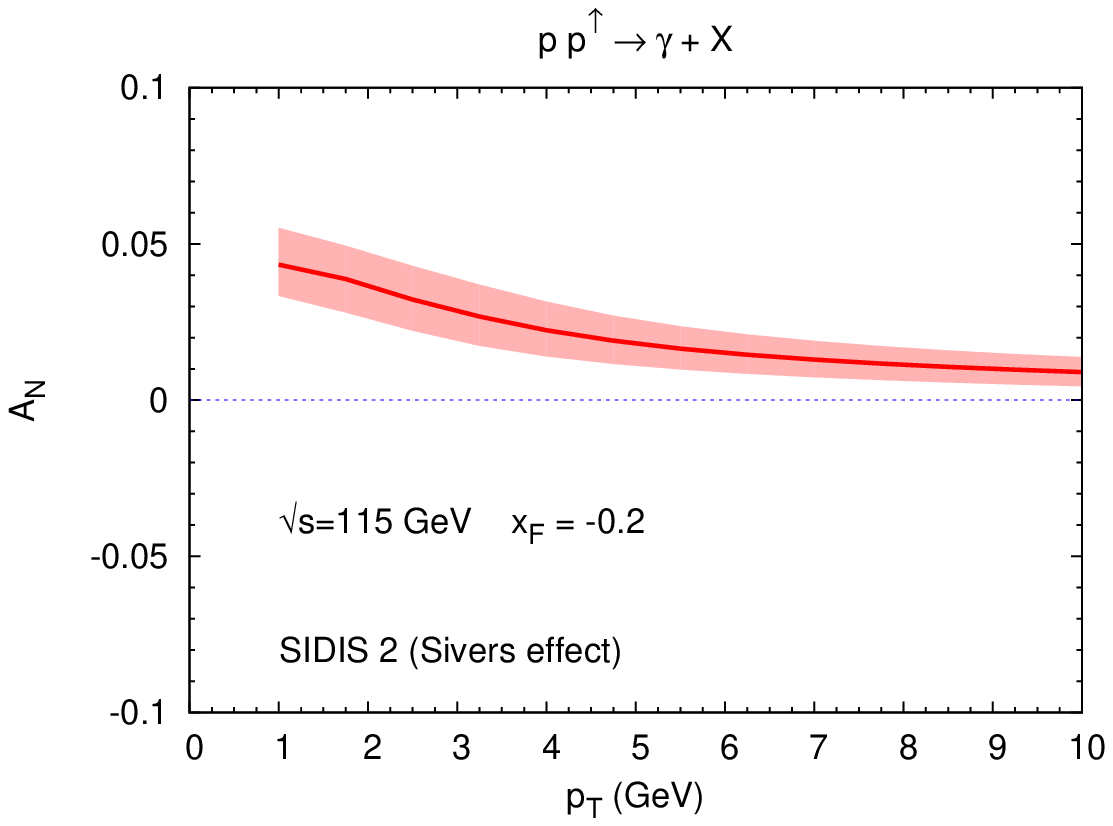}
\vskip 0.15 truecm
\includegraphics[width=8.5truecm,angle=0]{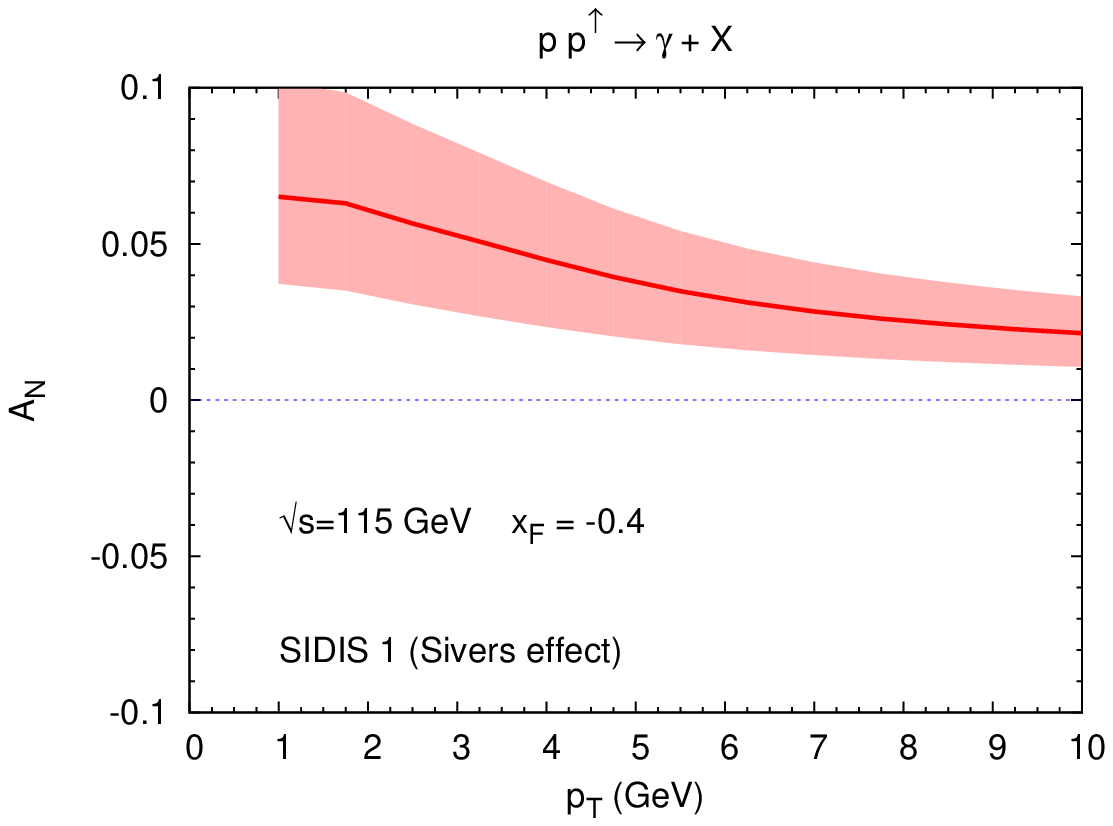}
\includegraphics[width=8.5truecm,angle=0]{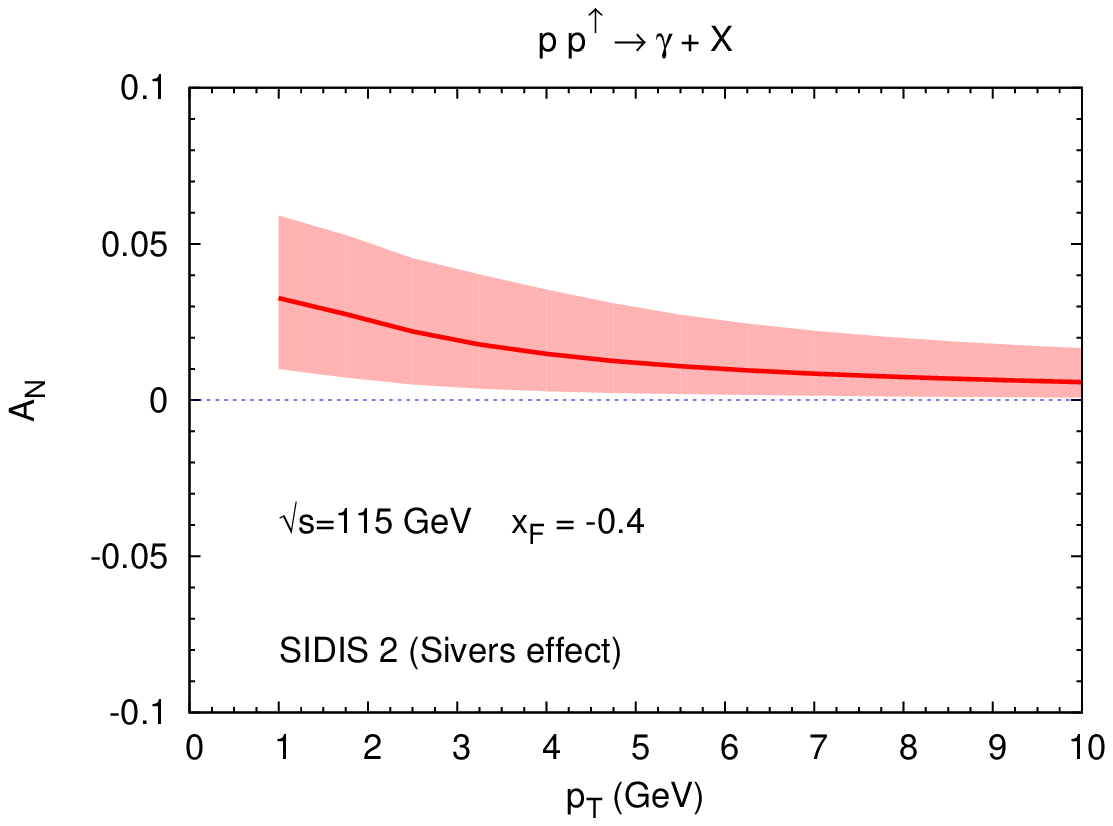}
\caption{Our theoretical estimates for $A_N$ vs.~$p_T$ at $\sqrt{s} = 115$ GeV,
$x_F = -0.2$ (upper plots) and $x_F = -0.4$ (lower plots) for inclusive photon
production in $p \, \pup \to \gamma \, X$ processes, computed
according to Eqs~(\ref{ansc}) and (\ref{numans}) of the text. Only the Sivers
effect contributes. The computation is performed adopting the Sivers functions
of Ref.~\cite{Anselmino:2005ea} (SIDIS 1, left panels) and of
Ref.~\cite{Anselmino:2008sga} (SIDIS 2, right panels). The overall statistical
uncertainty band, also shown, is obtained following the procedure described in
Appendix A of Ref.~\cite{Anselmino:2008sga}.}
\label{fig4}
\end{figure}

\begin{figure}[h!t]
\includegraphics[width=8.5truecm,angle=0]{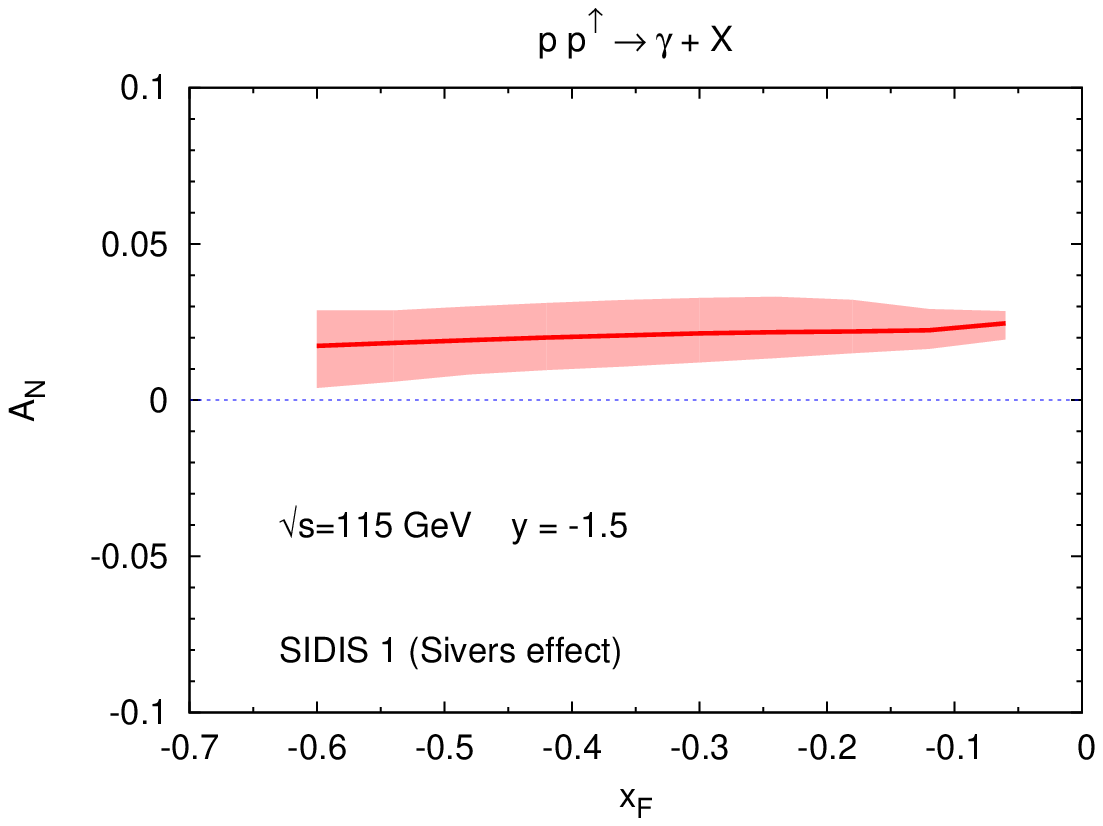}
\includegraphics[width=8.5truecm,angle=0]{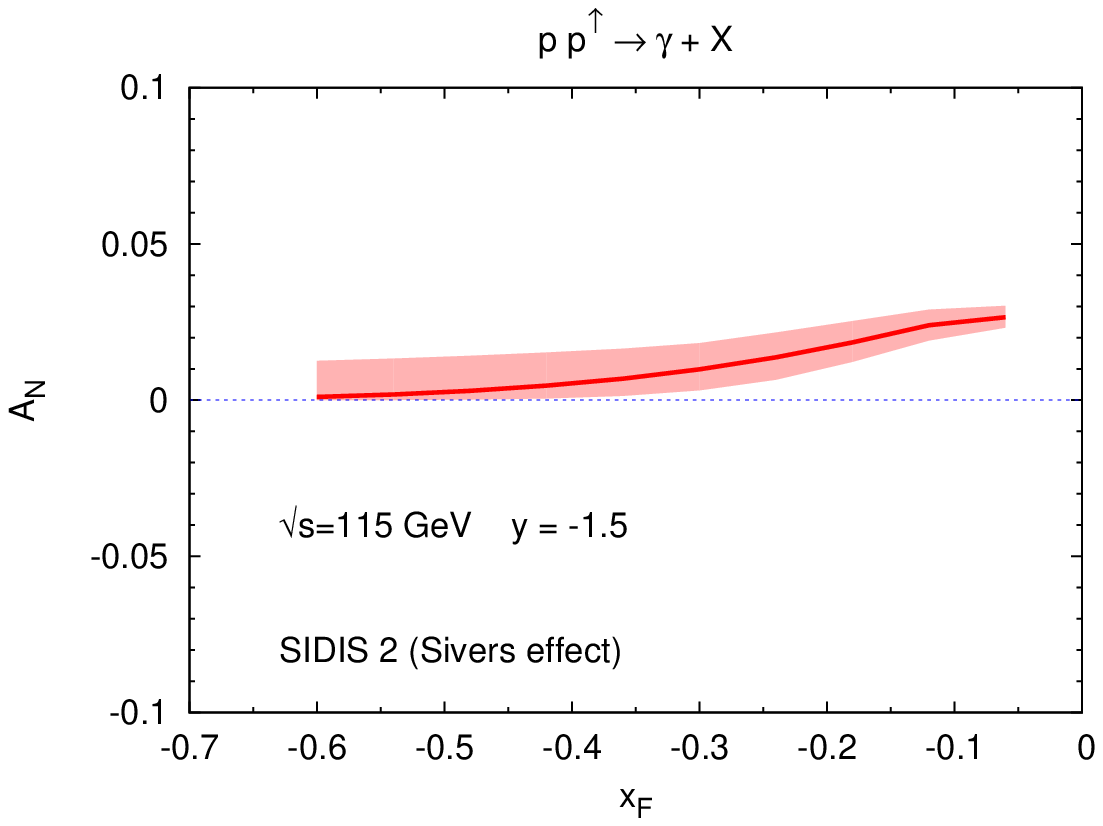}
\vskip 0.15 truecm
\includegraphics[width=8.5truecm,angle=0]{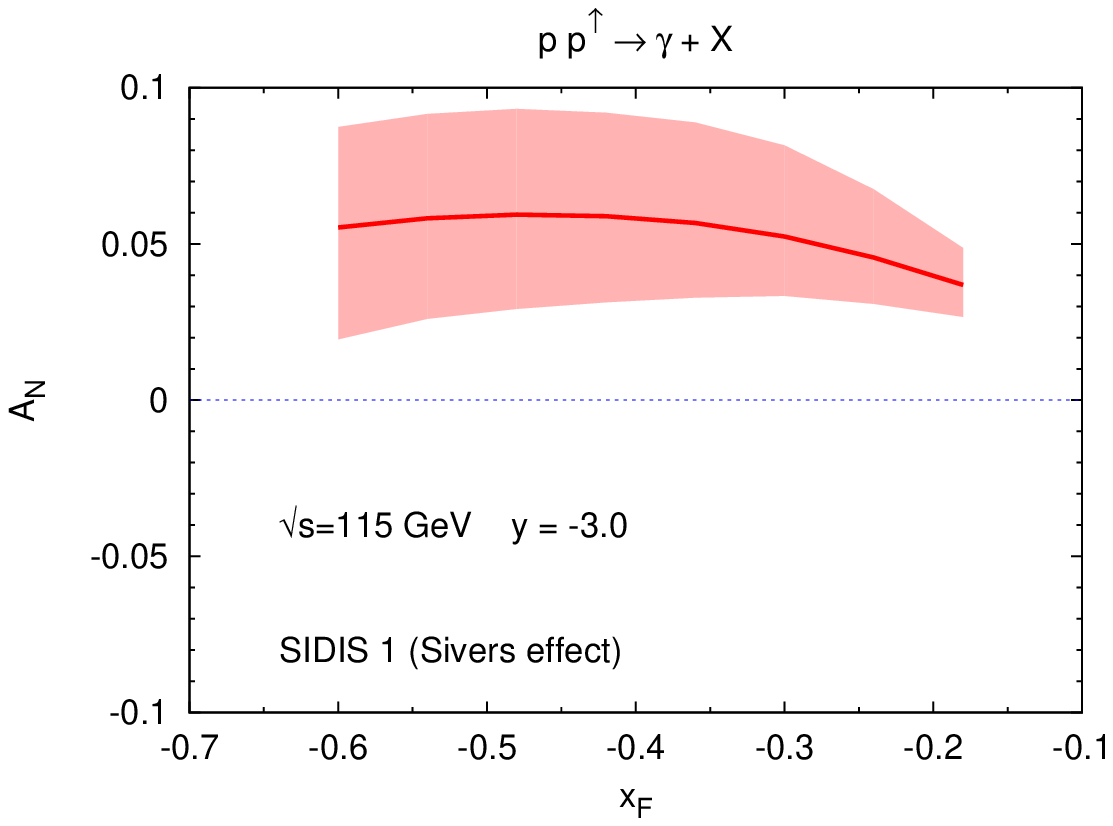}
\includegraphics[width=8.5truecm,angle=0]{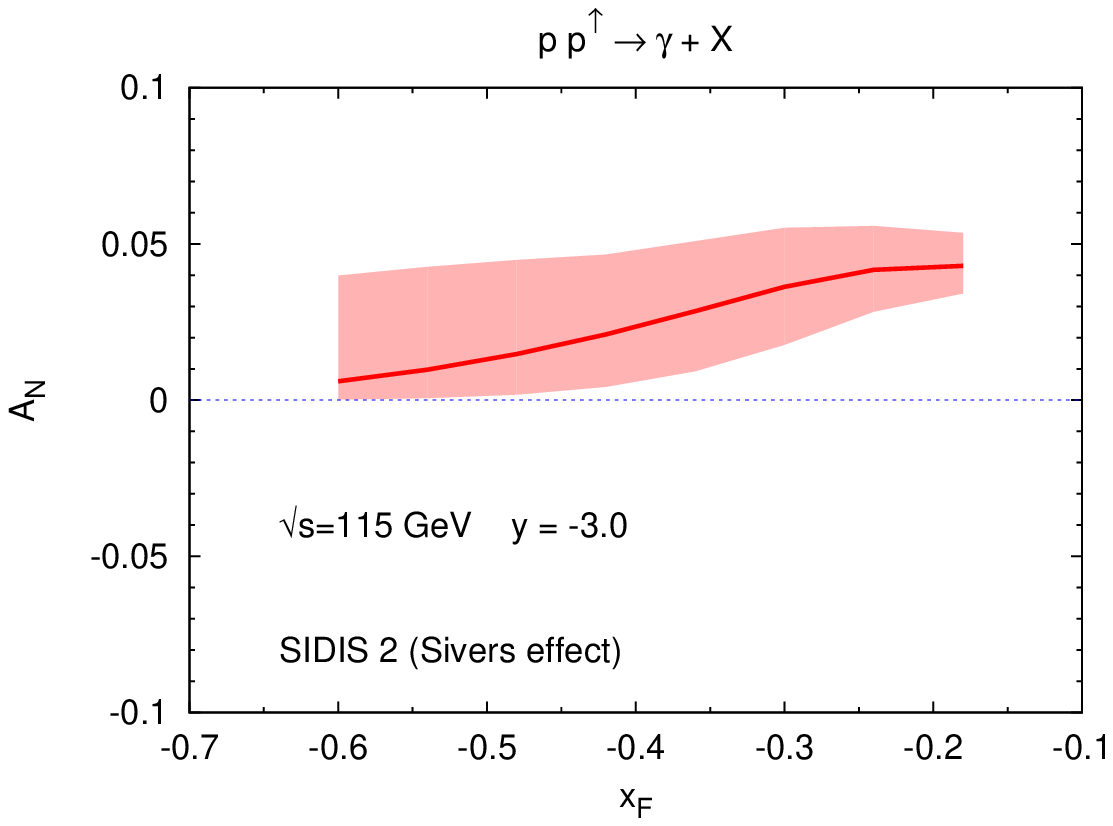}
\caption{Our theoretical estimates for $A_N$ vs.~$x_F$ at $\sqrt{s} = 115$ GeV,
$y = -1.5$ (upper plots) and $y = -3.0$ (lower plots) for inclusive photon
production in $p \, \pup \to \gamma \, X$ processes, computed
according to Eqs~(\ref{ansc}) and (\ref{numans}) of the text. Only the Sivers
effect contributes. The computation is performed adopting the Sivers functions
of Ref.~\cite{Anselmino:2005ea} (SIDIS 1, left panels) and of
Ref.~\cite{Anselmino:2008sga} (SIDIS 2, right panels). The overall statistical
uncertainty band, also shown, is obtained following the procedure described in
Appendix A of Ref.~\cite{Anselmino:2008sga}.}
\label{fig5}
\end{figure}

The analogous results for the single direct photon are shown in
Figs.~\ref{fig4}-\ref{fig6} (where $x_F =
2(\bfp_{\rm jet})_z/\sqrt s$), and those for the single jet production in
Figs.~\ref{fig7}-\ref{fig9} ($x_F= 2(\bfp_{\gamma})_z/\sqrt s$).
In these cases, obviously, there is no fragmentation process and only the Sivers
effect contributes to $A_N$, with $D_{h/c}(z, p_\perp)$ simply replaced by
$\delta(z-1)\,\delta^2(\bfp_\perp)$ in Eq.~(\ref{numans}) (see
Ref.~\cite{Anselmino:2013rya} for further details).
In our leading order treatment the jet coincides with a single
final parton. Notice that for a jet production we have all the same QCD
subprocesses which contribute to hadron production, while for a direct photon
production the basic partonic subprocesses are the Compton scattering
$g\,q(\bar q) \to \gamma \, q(\bar q)$ and the annihilation process
$q\, \bar q \to \gamma \, g$~\cite{D'Alesio:2004up}.

\begin{figure}[h!t]
\includegraphics[width=8.2truecm,angle=0]{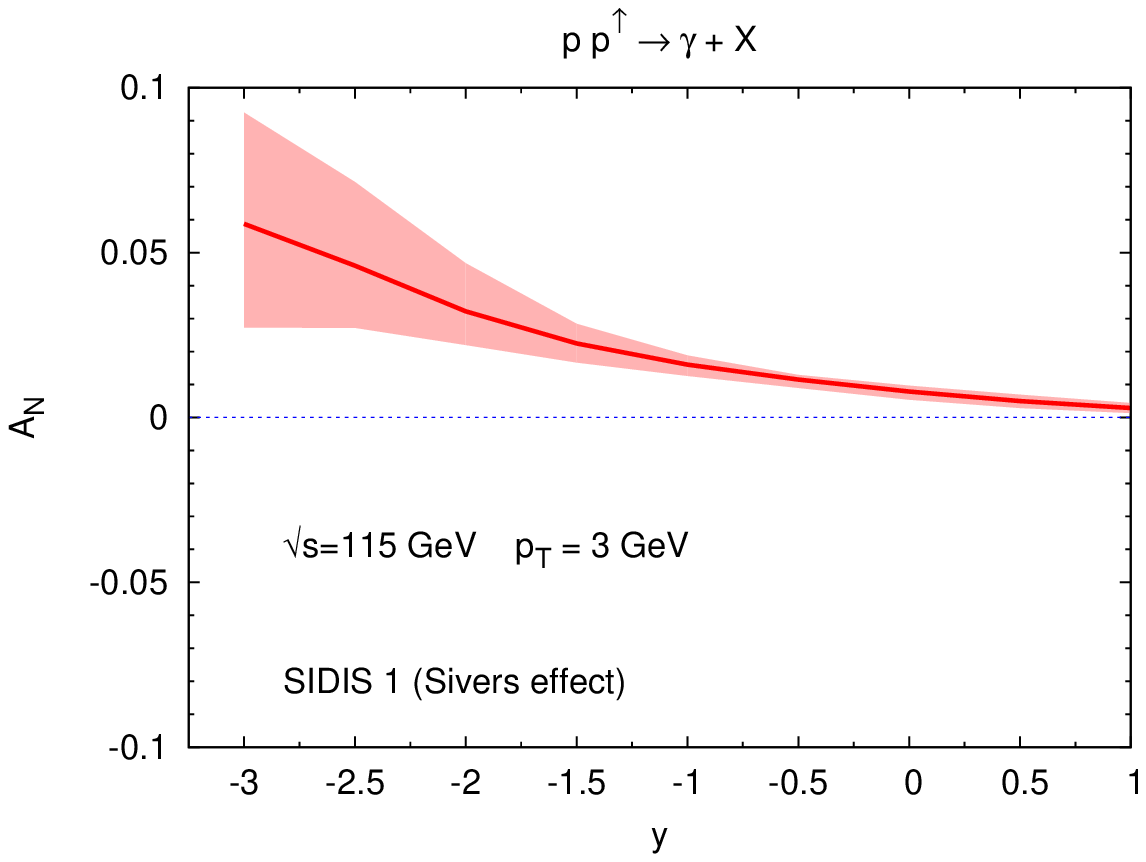}
\includegraphics[width=8.2truecm,angle=0]{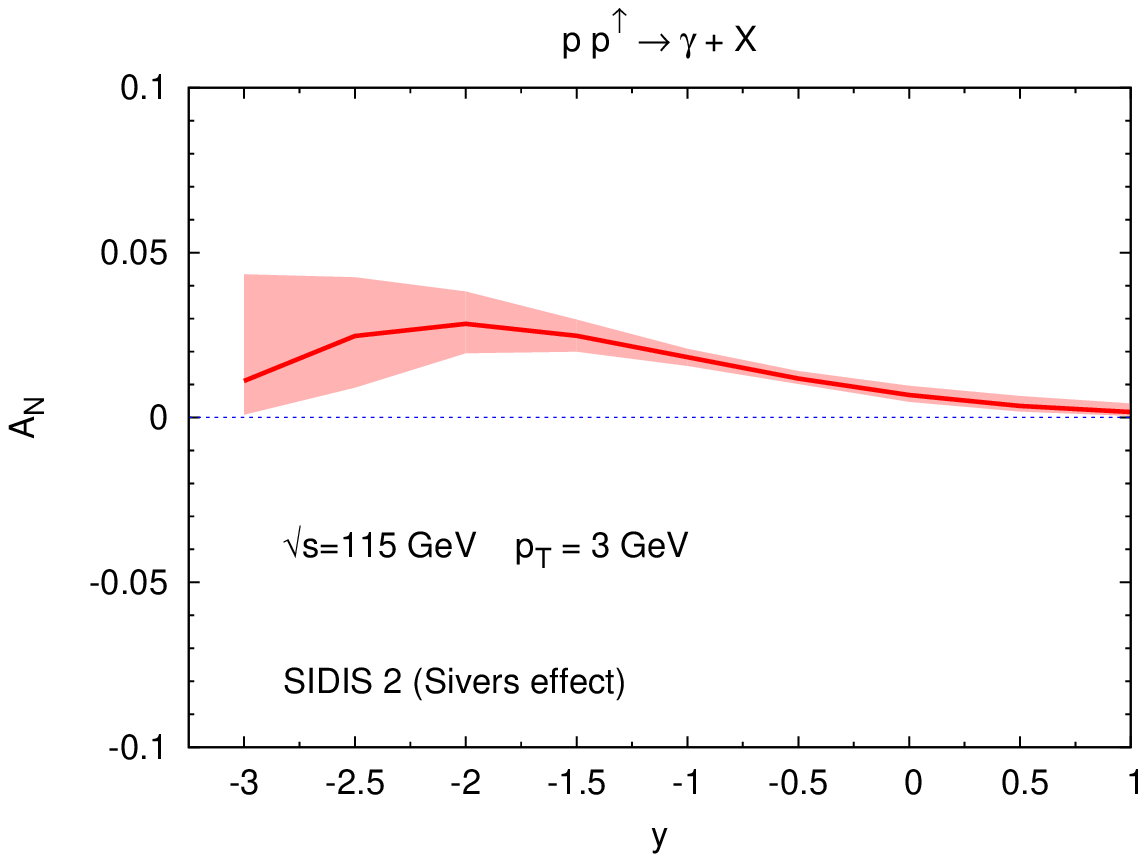}
\caption{Our theoretical estimates for $A_N$ vs.~$y$ at $\sqrt{s} = 115$ GeV
and $p_T = 3$ GeV, for inclusive photon production in $p \, \pup \to \gamma \, X$
processes, computed according to Eqs~(\ref{ansc})--(\ref{numanc}) of the text.
Only the Sivers effects contributes. The computation is performed adopting
the Sivers functions of Ref.~\cite{Anselmino:2005ea} (SIDIS 1, left panel)
and of Ref.~\cite{Anselmino:2008sga} (SIDIS 2, right panel). The overall
statistical uncertainty band, also shown, is the envelope
of the two independent statistical uncertainty bands obtained following the
procedure described in Appendix A of Ref.~\cite{Anselmino:2008sga}.}
\label{fig6}
\end{figure}
\begin{figure}[h!t]
\includegraphics[width=8.5truecm,angle=0]{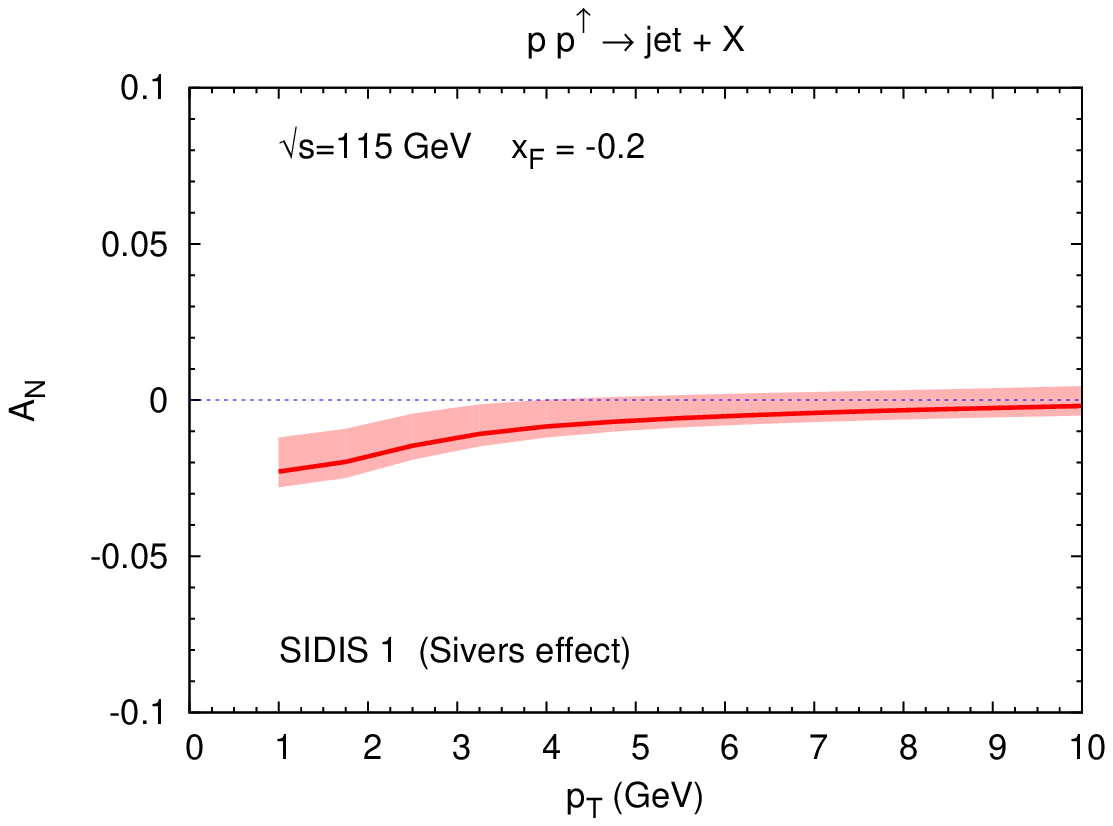}
\includegraphics[width=8.5truecm,angle=0]{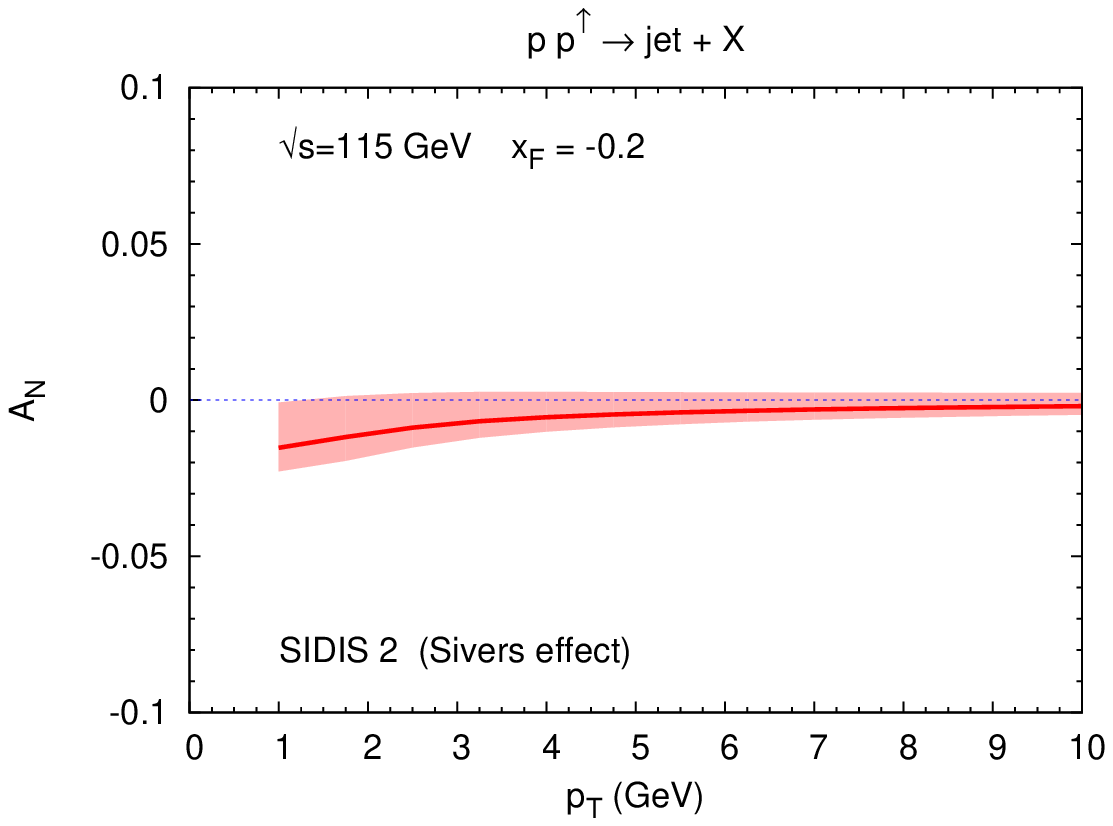}
\vskip 0.1 truecm
\includegraphics[width=8.5truecm,angle=0]{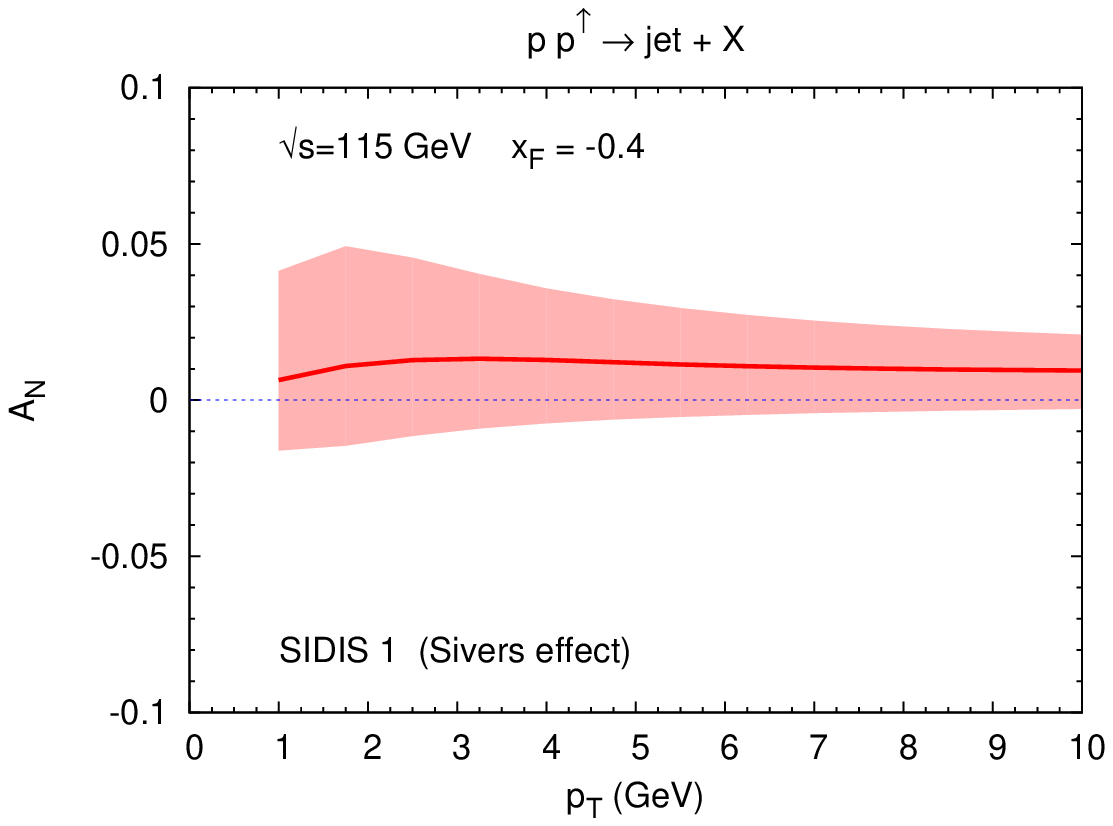}
\includegraphics[width=8.5truecm,angle=0]{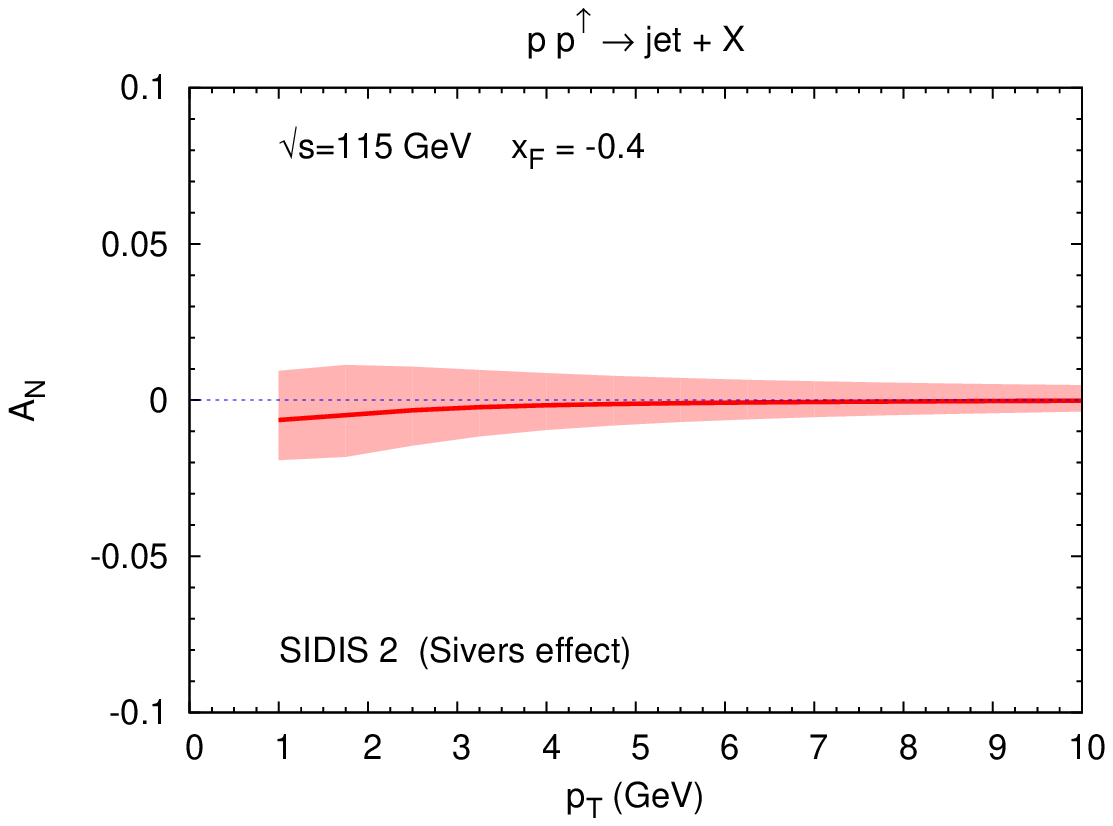}
\caption{Our theoretical estimates for $A_N$ vs.~$x_F$ at $\sqrt{s} = 115$ GeV,
$x_F = -0.2$ (upper plots) and $x_F = -0.4$ (lower plots) for inclusive single
jet production in $p \, \pup \to {\rm jet} \, X$ processes, computed
according to Eqs~(\ref{ansc}) and (\ref{numans}) of the text. Only the Sivers
effect contributes. The computation is performed adopting the Sivers functions
of Ref.~\cite{Anselmino:2005ea} (SIDIS 1, left panels) and of
Ref.~\cite{Anselmino:2008sga} (SIDIS 2, right panels). The overall statistical
uncertainty band, also shown, is obtained following the procedure described in
Appendix A of Ref.~\cite{Anselmino:2008sga}.}
\label{fig7}
\end{figure}

\section{$A_N$ for Drell-Yan processes}

Drell-Yan (D-Y) processes are expected to play a crucial role in our
understanding of the origin, at the partonic level, of TSSAs. For such
processes, like for SIDIS processes and contrary to single hadron production,
the TMD factorisation has been proven to hold, so that there is a general
consensus that the Sivers effect should be visible via TSSAs in D-Y \cite{Collins:1984kg, Ji:2004xq, Collins:2011zzd, GarciaEchevarria:2011rb}. Not only: the widely accepted interpretation of the QCD origin of TSSAs as final
or initial state interactions of the scattering partons~\cite{Brodsky:2002cx}
leads to the conclusion that the Sivers function has opposite signs in SIDIS
and D-Y processes~\cite{Collins:2002kn}. Which remains to be seen.

Predictions for Sivers $A_N$ in D-Y and at different possible experiments
were given in Ref.~\cite{Anselmino:2009st}, which we follow here.

In Ref.~\cite{Anselmino:2009st} predictions were given for the $\pup p \to
\ell^+\ell^- X$ D-Y process in the $\pup-p$ c.m. frame, in which one observes
the four-momentum $q$ of the final $\ell^+\ell^-$ pair. Notice that $q^2 = M^2$
is the large scale in the process, while $q_T = |\bfq_T|$ is the small one.
In order to collect data at all azimuthal angles, one defines the weighted
spin asymmetry:
\bea
&& A_N^{\sin(\phi_{\gamma} - \phi_S)} \equiv
\frac{\int_0^{2\pi} d\phi_{\gamma} \>
[d\sigma^{\uparrow} - d\sigma^{\downarrow}] \>
\sin(\phi_{\gamma}-\phi_S)}
{\frac{1}{2}\int_{0}^{2\pi} d\phi_{\gamma} \>
[d\sigma^{\uparrow} + d\sigma^{\downarrow}]} \label{ANW} \\
&=& \frac{\int d\phi_{\gamma} \> \left[
\sum_q e_q^2 \int d^2\bfk_{\perp 1} \, d^2\bfk_{\perp 2} \>
\delta^2(\bfk_{\perp 1} + \bfk_{\perp 2} - \bfq_T) \>
\Delta^N\!f_{q/\pup}(x_1, \bfk_{\perp 1}) \>
 f_{\bar q/p}(x_{2}, k_{\perp 2}) \right] \> \sin(\phi_{\gamma}-\phi_S)}
{\int d\phi_{\gamma} \> \left[
\sum_q e_q^2 \int d^2\bfk_{\perp 1} \, d^2\bfk_{\perp 2} \>
\delta^2(\bfk_{\perp 1} + \bfk_{\perp 2} - \bfq_T) \>
 f_{q/p}(x_1, k_{\perp 1}) \>
 f_{\bar q/p}(x_{2}, k_{\perp 2})\right]}
 \> , \label{ANW2}
 \eea
where $\phi_\gamma$ and $\phi_S$ are respectively the azimuthal angle of the
$\ell^+\ell^-$ pair and of the proton transverse spin and we have defined (see
Eq.~(\ref{sivnoi})):
\be
\Delta^N\!f_{q/\pup}(x, \bfk_{\perp}) \equiv
\Delta^N\!f_{q/\pup}(x, k_{\perp}) \>
\bfS \cdot (\hat{\bfP} \times \hat{\bfk}_{\perp}) =
\hat f_{q/\pup}(x, \bfk_{\perp}) - \hat f_{q/\pdown}(x, \bfk_{\perp}) \>.
\label{f-f}
\ee

Adopting for the unpolarised TMD and the Sivers function the same expressions
as in Eqs.~(\ref{TMDpdf}) and (\ref{eq:siv-par})-(\ref{eq:h-siv}) allows,
at {$\cal O$}($k_\perp/M$), an analytical integration of the numerator and
denominator of Eq.~(\ref{ANW2}), resulting in a simple expression for the
asymmetry $A_N^{\sin(\phi_{\gamma} - \phi_S)}$~\cite{Anselmino:2009st}.

Notice that we consider here the $p \, \pup \to \ell^+\ell^- X$ D-Y process
in the $p-\pup$ c.m. frame. For such a process the TSSA is given
by~\cite{Anselmino:2009st}
\be
A_N^{\sin(\phi_{\gamma} - \phi_S)}(p \, \pup \to \gamma^* X;
\> x_F, M, q_T) = - A_N^{\sin(\phi_{\gamma} - \phi_S)}(\pup \,
p \to \gamma^* X; \> -x_F, M, q_T) \>. \label{ab-ba}
\ee

Our results for the Sivers asymmetry $A_N^{\sin(\phi_{\gamma} - \phi_S)}$
at AFTER@LHC, obtained following Ref.~\cite{Anselmino:2009st},
Eq.~(\ref{ab-ba}) and using the SIDIS extracted Sivers function
{\it reversed in sign}, are shown in Fig.~\ref{fig10}. Further details can be
found in the captions of these figures.

\section{Comments and conclusions}

Some final comments and further details might help in understanding the importance of
the measurements of the TSSAs at AFTER@LHC.
\begin{itemize}
\item
 Most predictions given show clear asymmetries, sufficiently
large as to be easily measurable, given the expected performance of
AFTER@LHC~\cite{Brodsky:2012vg}. The uncertainty
bands reflects the uncertainty in the extraction of the Sivers and
transversity functions from SIDIS data, which are focused on small and
intermediate $x$ values ($x \lsim 0.3$); in fact the bands grow larger at
larger values of $|x_F|$.
\item
The values of $A_N$ found for pion production can be as large as 10\% for
$\pi^\pm$, while they are smaller for $\pi^0$. They result from the sum of
the Sivers and the Collins effects. The relative importance of the two
contributions varies according to the kinematical regions and the set of
distributions and fragmentation functions adopted. As a tendency, the
contribution from the Sivers effect is larger than the Collins contribution
with the SIDIS 1 - KRE set, while the opposite is true for the SIDIS 2 - DSS
set.

The values found here are in agreement, both in sign and qualitative magnitude,
with the values found in Ref.~\cite{Kanazawa:2015fia} within the collinear
twist-3 (CT-3) approach.
\item
The results for single photon production are interesting; they isolate the
Sivers effect and our predictions show that they can reach values of about
5\%, with a reduced uncertainty band. We find positive values of $A_N$ as
the relative weight of the quark charges leads to a dominance of the $u$
quark and the Sivers functions $\Delta^N \! f_ {u/\pup}$ is
positive~\cite{Anselmino:2005ea, Anselmino:2008sga}.

Our results, obtained within the GPM, have a similar magnitude as those
obtained in Refs.~\cite{Kanazawa:2015fia} and~\cite{Kanazawa:2014nea},
within the CT-3 approach, but have {\it an opposite sign}. Thus, a
measurement of $A_N$ for a single photon production, although difficult,
would clearly discriminate between the two approaches.
\item
The values of $A_N$ for single jet production, which might be interesting
as they also have no contribution from the Collins effect, turn out to be
very small and compatible with zero, due to a strong cancellation between
the $u$ and $d$ quark contributions. The same result is found in
Ref.~\cite{Kanazawa:2015fia}.
\item
A measurement of $A_N^{\sin(\phi_{\gamma} - \phi_S)}$ in D-Y processes at
AFTER@LHC is a most interesting one. In such a case the TMD factorisation
 has been shown to be valid
and the Sivers asymmetry should show the expected sign
change with respect to SIDIS processes~\cite{Brodsky:2002cx,Collins:2002kn}.
Our computations, Fig.~\ref{fig10}, predict a clear asymmetry, which can be
as sizeable as 10\%, with a definite sign, even within the uncertainty band.
\end{itemize}

Both the results of Ref.~\cite{Kanazawa:2015fia} and the results of this
paper, obtain solid non negligible values for the TSSA $A_N$ measurable at
the AFTER@LHC experiment. The two sets of results are based on different
approaches, respectively the CT-3 and the GPM factorisation schemes.
While the magnitude of $A_N$ is very similar in the two cases, the
signs can be different; in particular, the TSSA for a direct photon
production, $p \, \pup \to \gamma \, X$, has opposite signs in the two
schemes.

In this paper we have also considered azimuthal asymmetries in polarised
D-Y processes, related to the Sivers effect. As explained above, in this
case, due to the presence of a large and a small scale, like in SIDIS, the
TMD factorisation is valid, with the expectation of an opposite sign of the
Sivers function in SIDIS and D-Y processes. Also this prediction can be
checked at AFTER@LHC.

\begin{figure}[h!t]
\includegraphics[width=8.5truecm,angle=0]{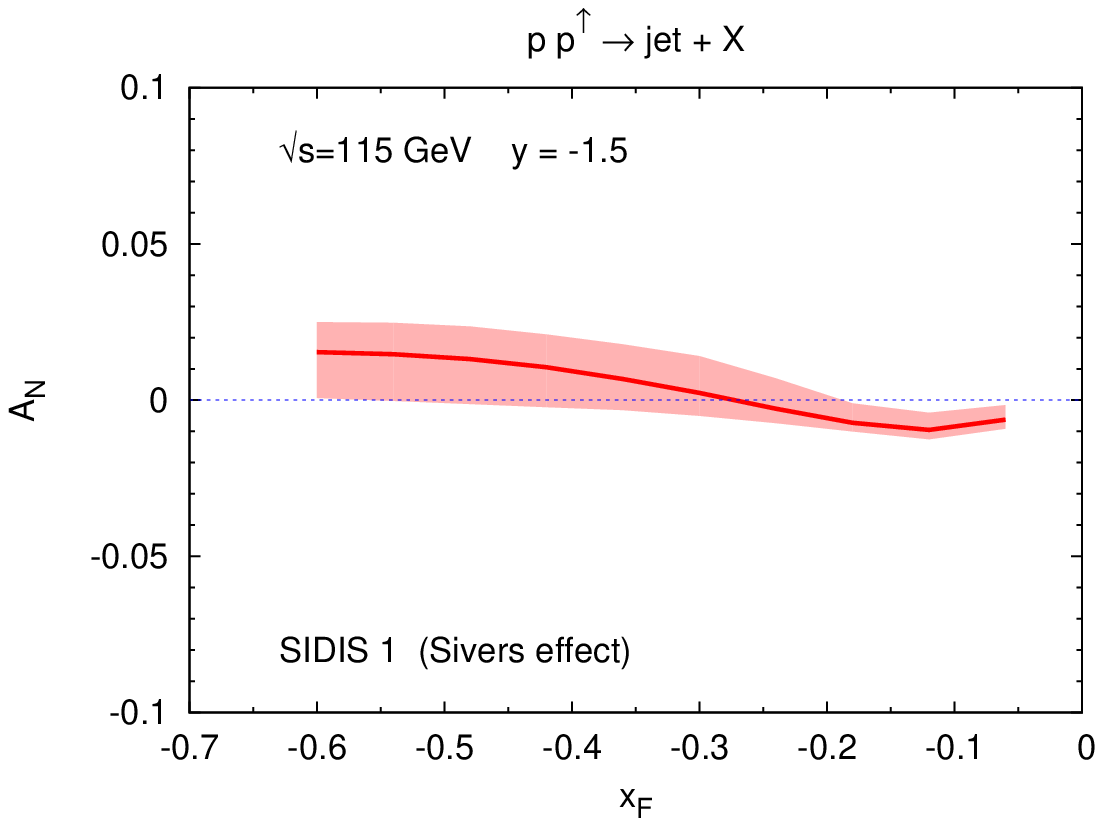}
\includegraphics[width=8.5truecm,angle=0]{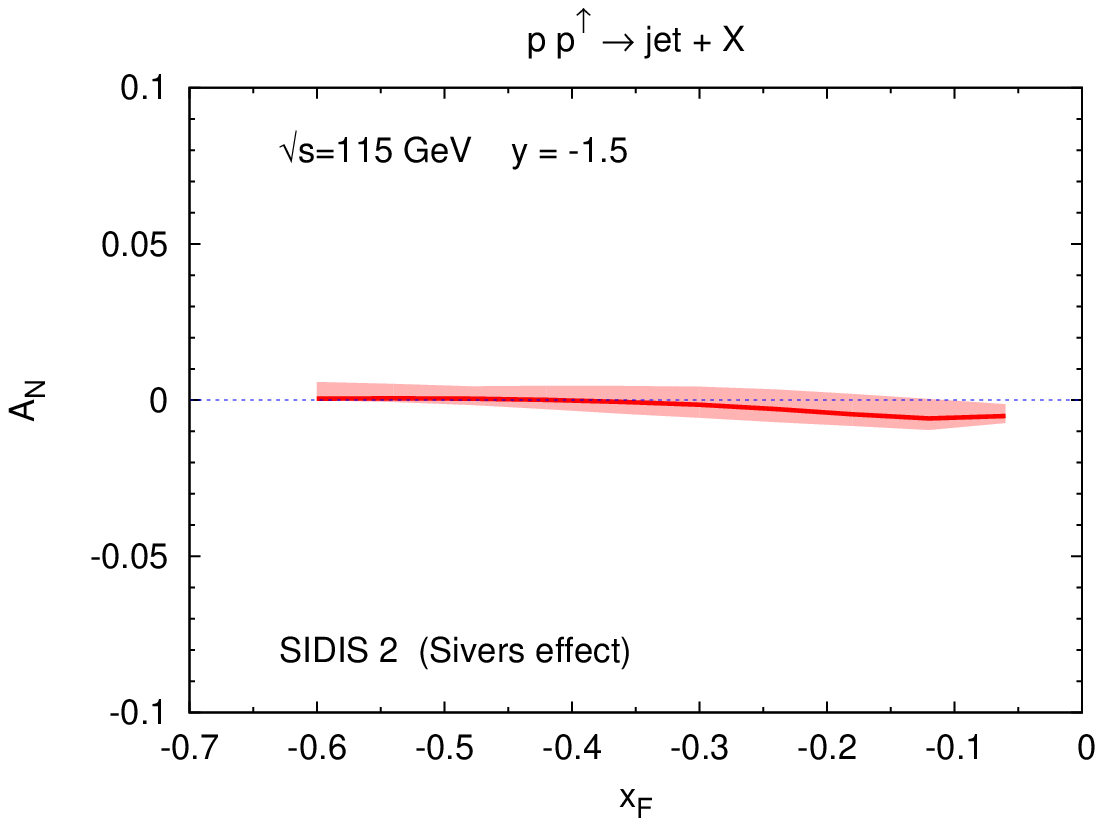}
\vskip 0.1 truecm
\includegraphics[width=8.5truecm,angle=0]{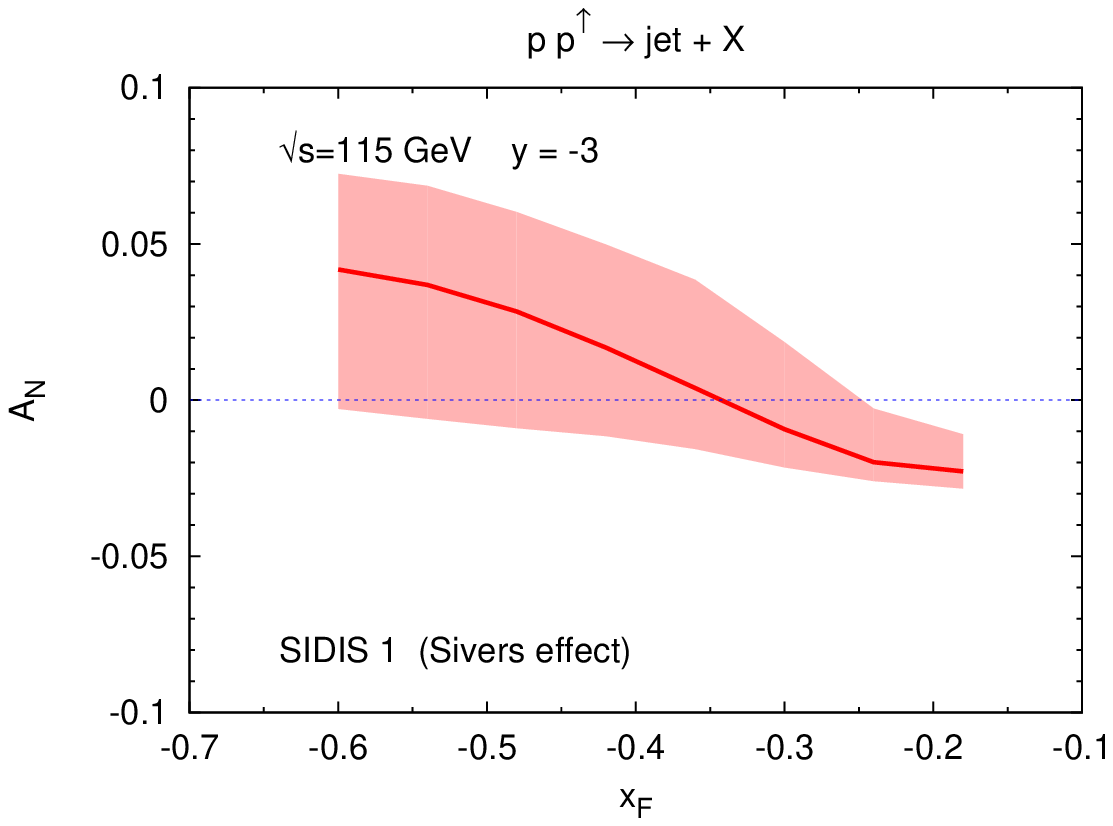}
\includegraphics[width=8.5truecm,angle=0]{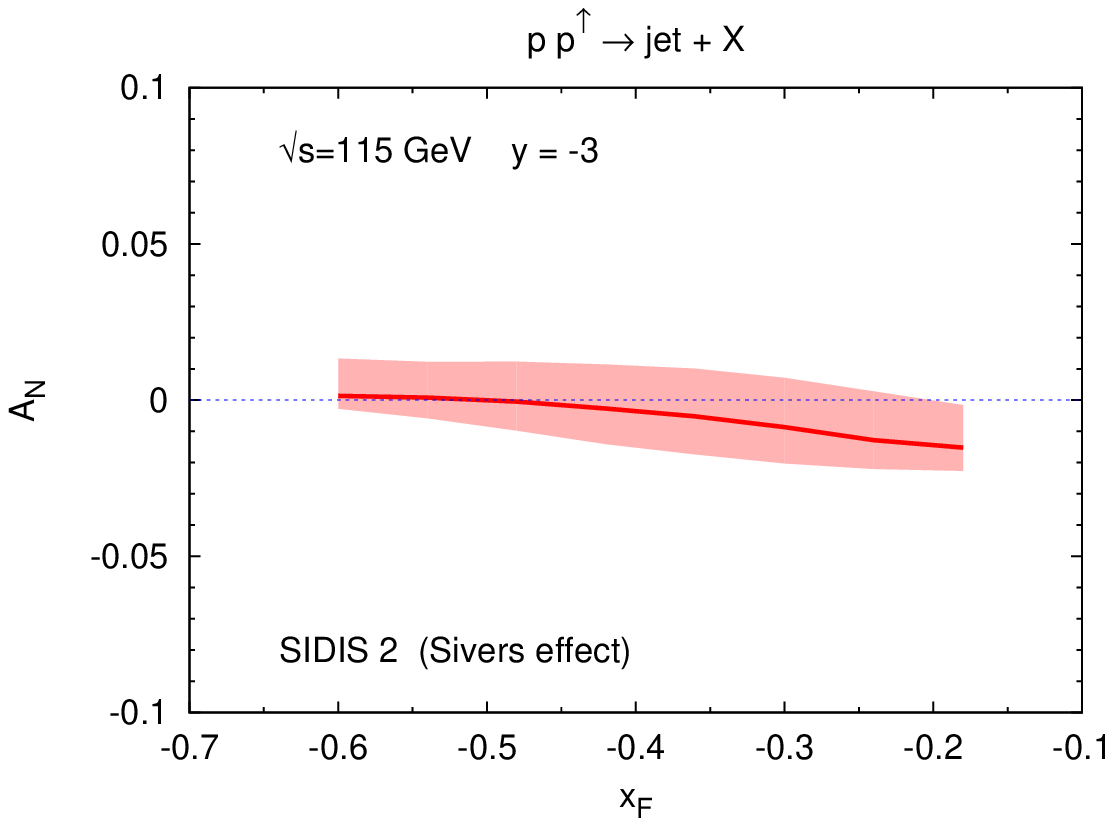}
\caption{Our theoretical estimates for $A_N$ vs.~$x_F$ at $\sqrt{s} = 115$ GeV,
$y = -1.5$ (upper plots) and $y = -3.0$ (lower plots) for inclusive single
jet production in $p \, \pup \to {\rm jet} \, X$ processes, computed
according to Eqs~(\ref{ansc}) and (\ref{numans}) of the text. Only the Sivers
effect contributes. The computation is performed adopting the Sivers functions
of Ref.~\cite{Anselmino:2005ea} (SIDIS 1, left panels) and of
Ref.~\cite{Anselmino:2008sga} (SIDIS 2, right panels). The overall statistical
uncertainty band, also shown, is obtained following the procedure described in
Appendix A of Ref.~\cite{Anselmino:2008sga}.}
\label{fig8}
\end{figure}
\begin{figure}[h!t]
\vspace*{-.2cm}
\includegraphics[width=8.5truecm,angle=0]{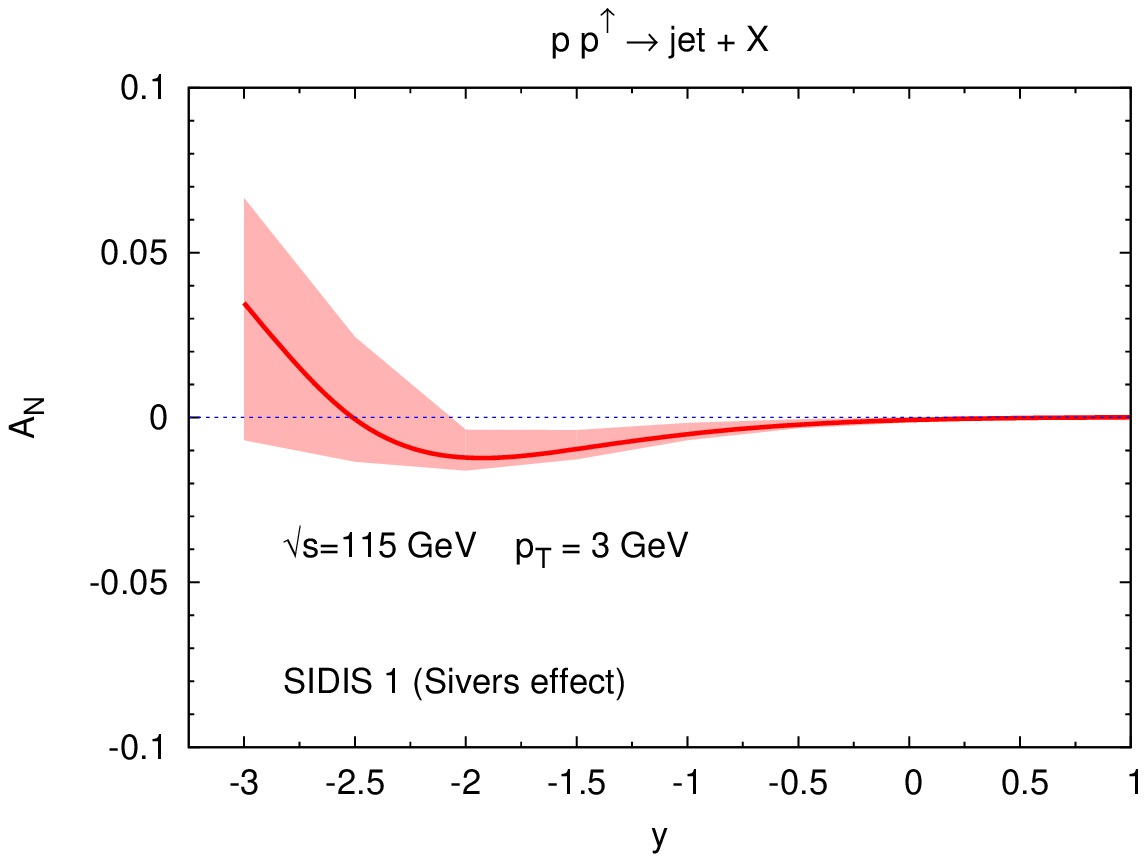}
\includegraphics[width=8.5truecm,angle=0]{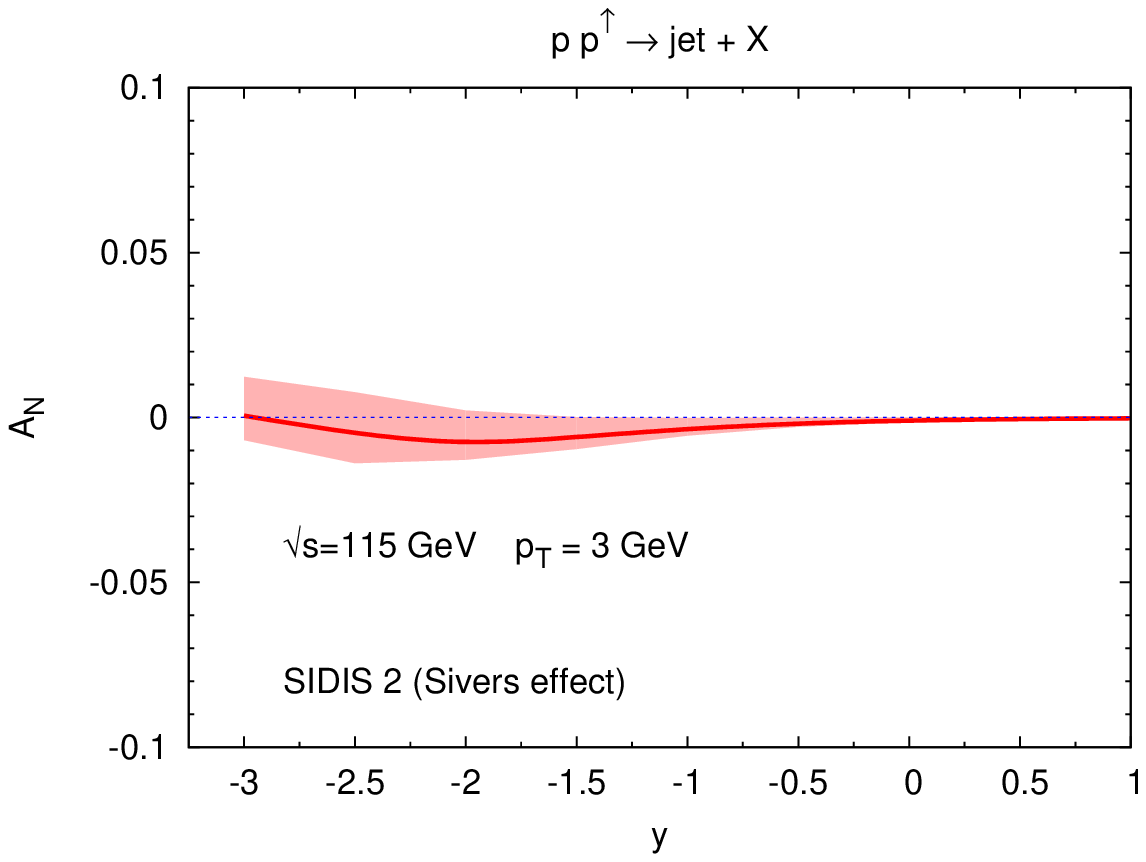}
\caption{Our theoretical estimates for $A_N$ vs.~$y$ at $\sqrt{s} = 115$ GeV
and $p_T = 3$ GeV, for inclusive single jet production in
$p \, \pup \to {\rm jet} \, X$ processes, computed according to
Eqs~(\ref{ansc})--(\ref{numanc}) of the text. Only the Sivers effect
contributes. The computation is performed adopting the Sivers
functions of Ref.~\cite{Anselmino:2005ea} (SIDIS 1, left panel) and
of Ref.~\cite{Anselmino:2008sga} (SIDIS 2, right panel). The overall
statistical uncertainty band, also shown, is the envelope of the two
independent statistical uncertainty bands obtained following the procedure
described in Appendix A of Ref.~\cite{Anselmino:2008sga}.}
\label{fig9}
\end{figure}
\begin{figure}[h!t]
\includegraphics[width=8.5truecm,angle=0]{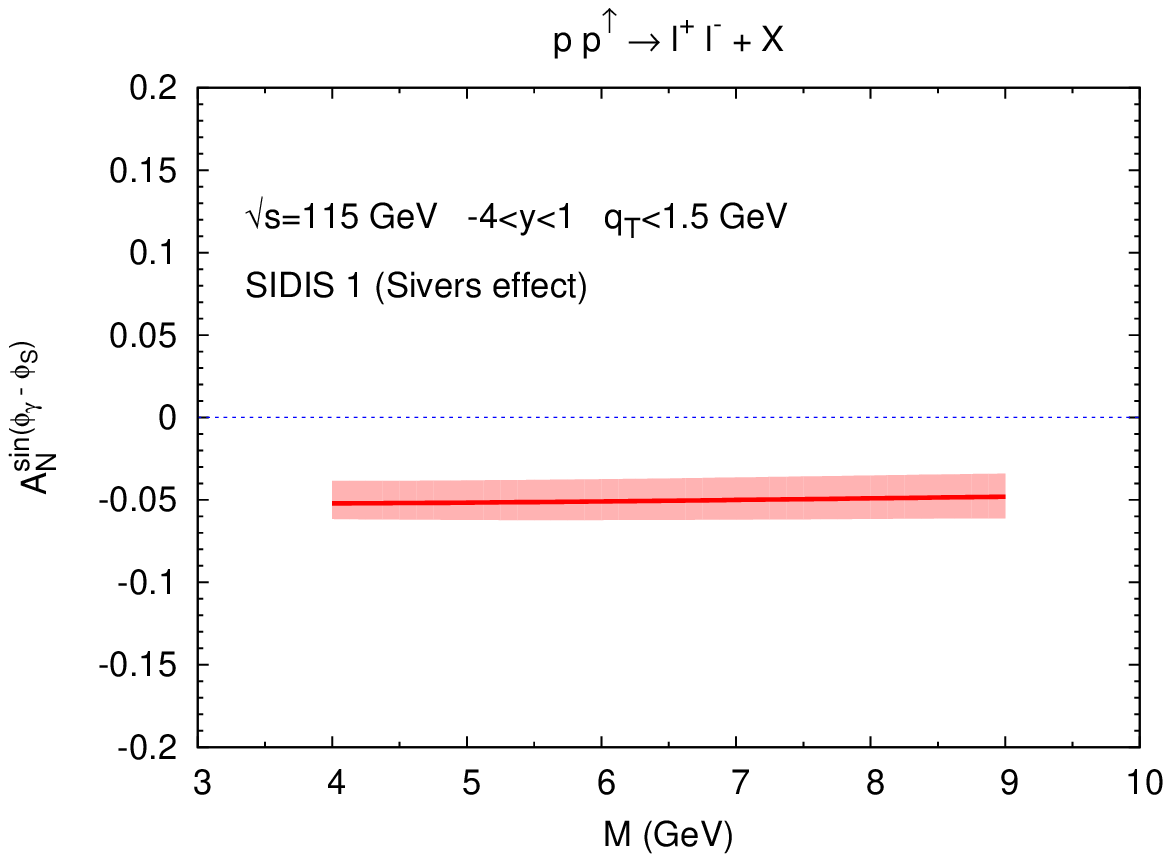}
\includegraphics[width=8.5truecm,angle=0]{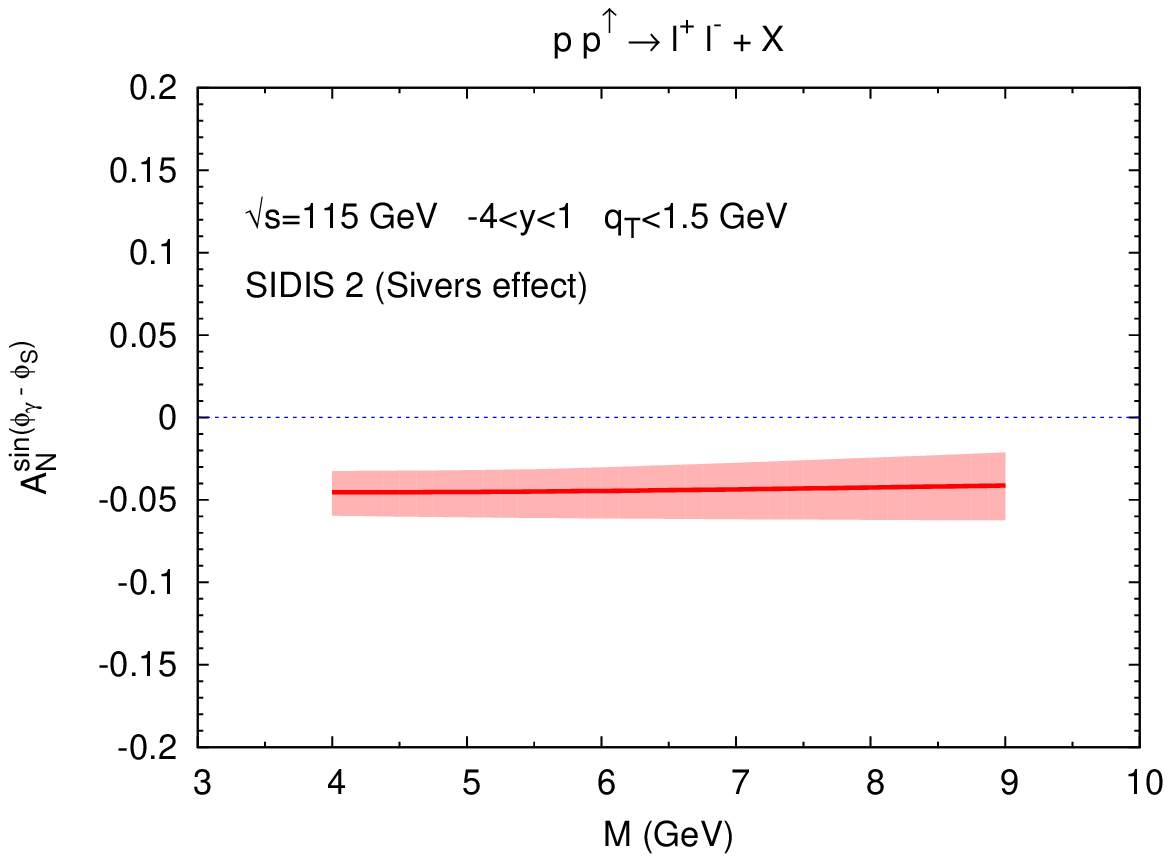}
\vskip 0.5 truecm
\includegraphics[width=8.5truecm,angle=0]{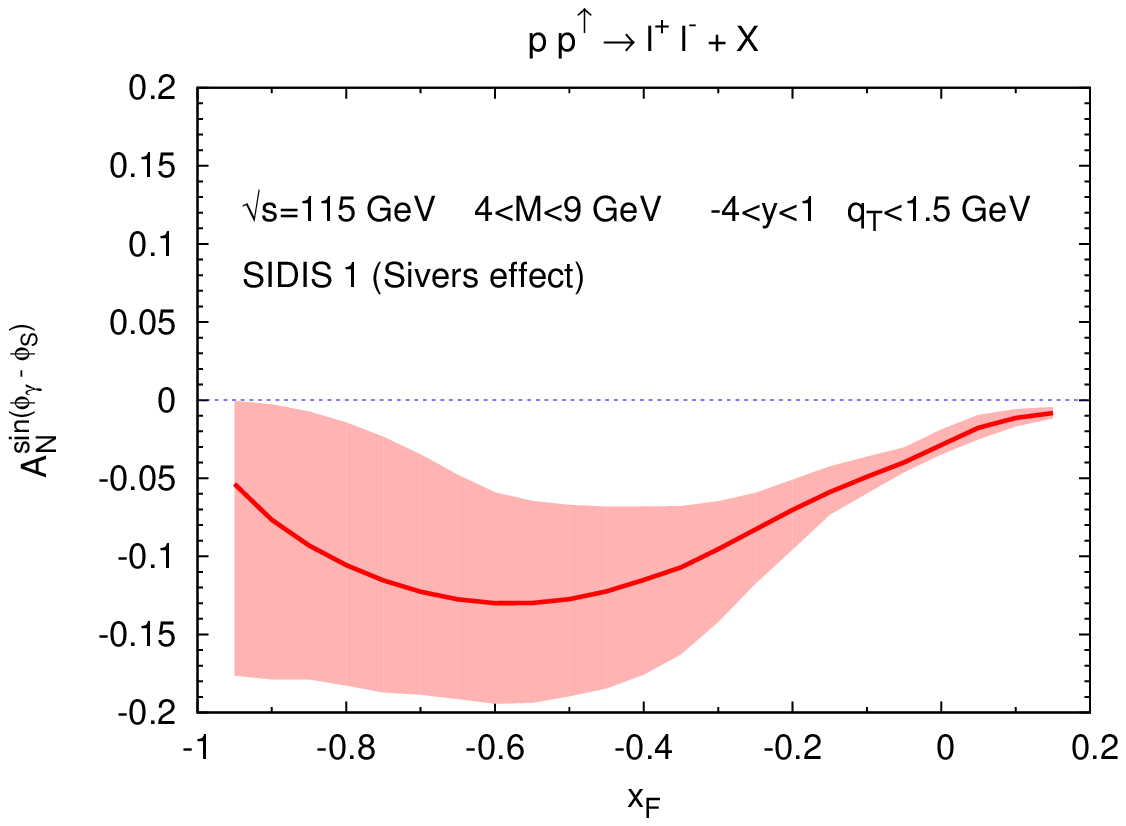}
\includegraphics[width=8.5truecm,angle=0]{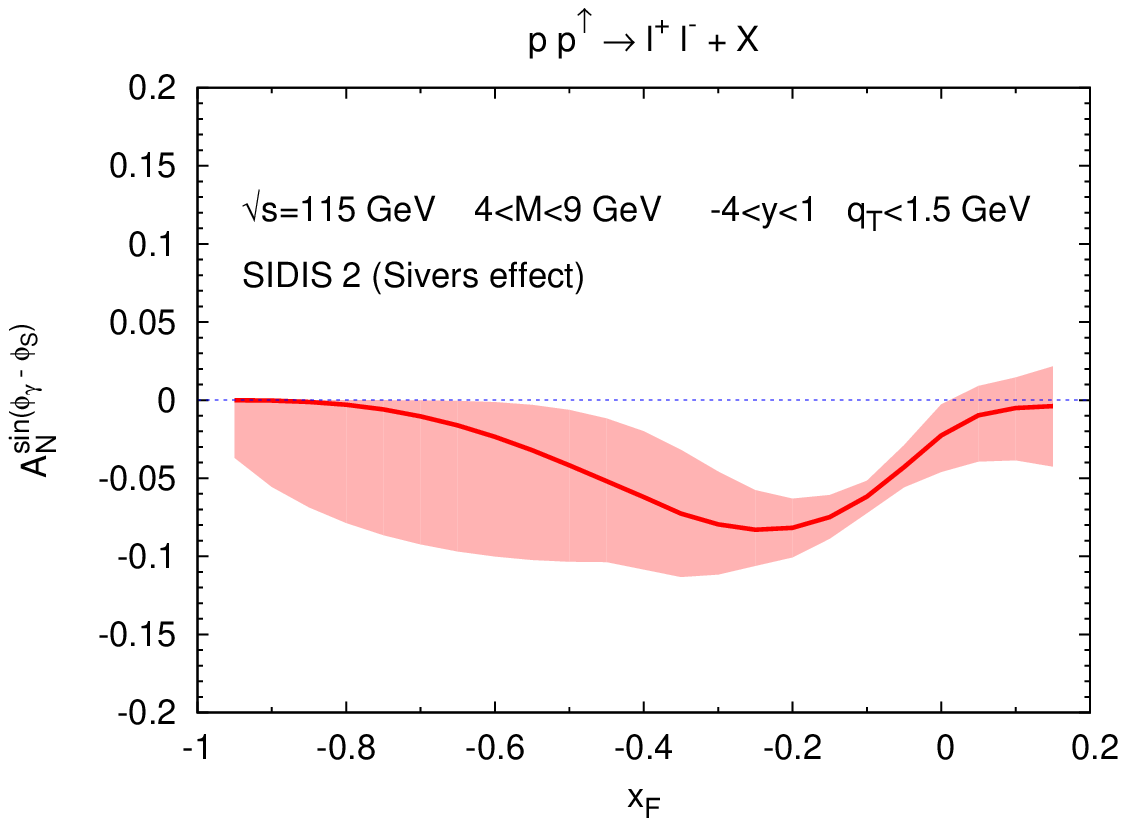}
\vskip 0.5 truecm
\includegraphics[width=8.5truecm,angle=0]{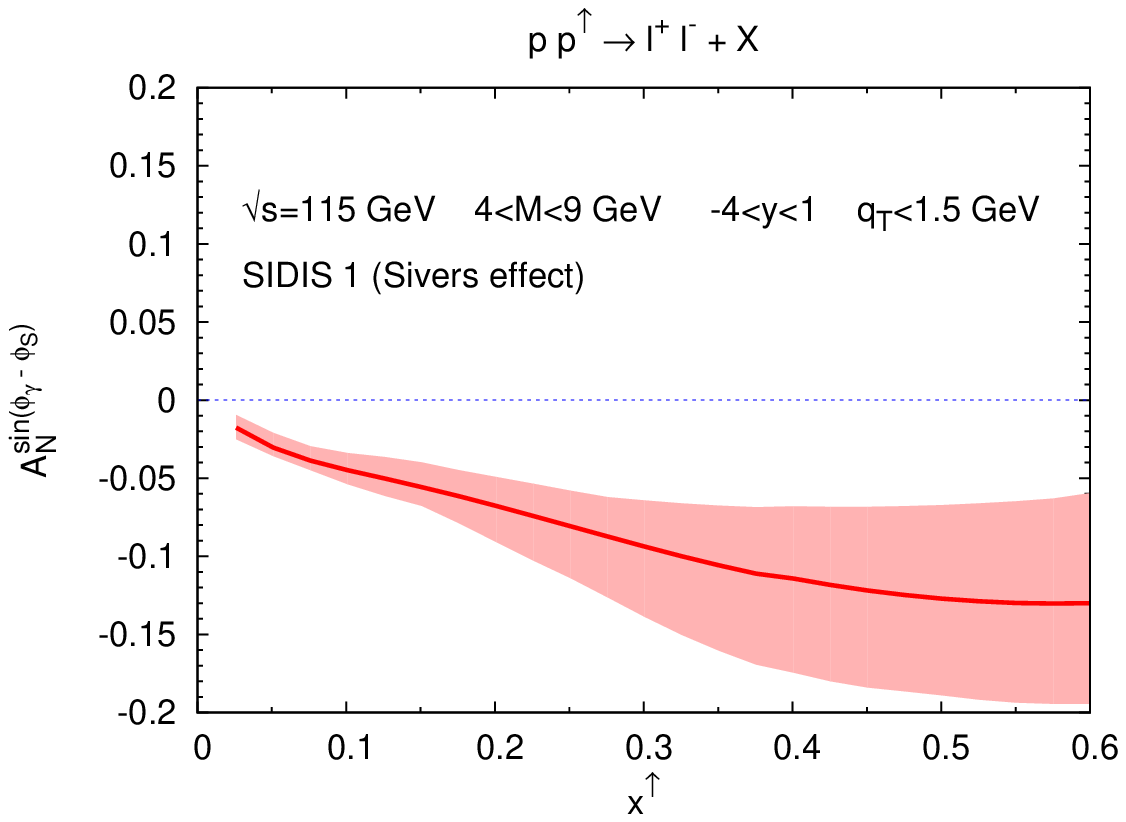}
\includegraphics[width=8.5truecm,angle=0]{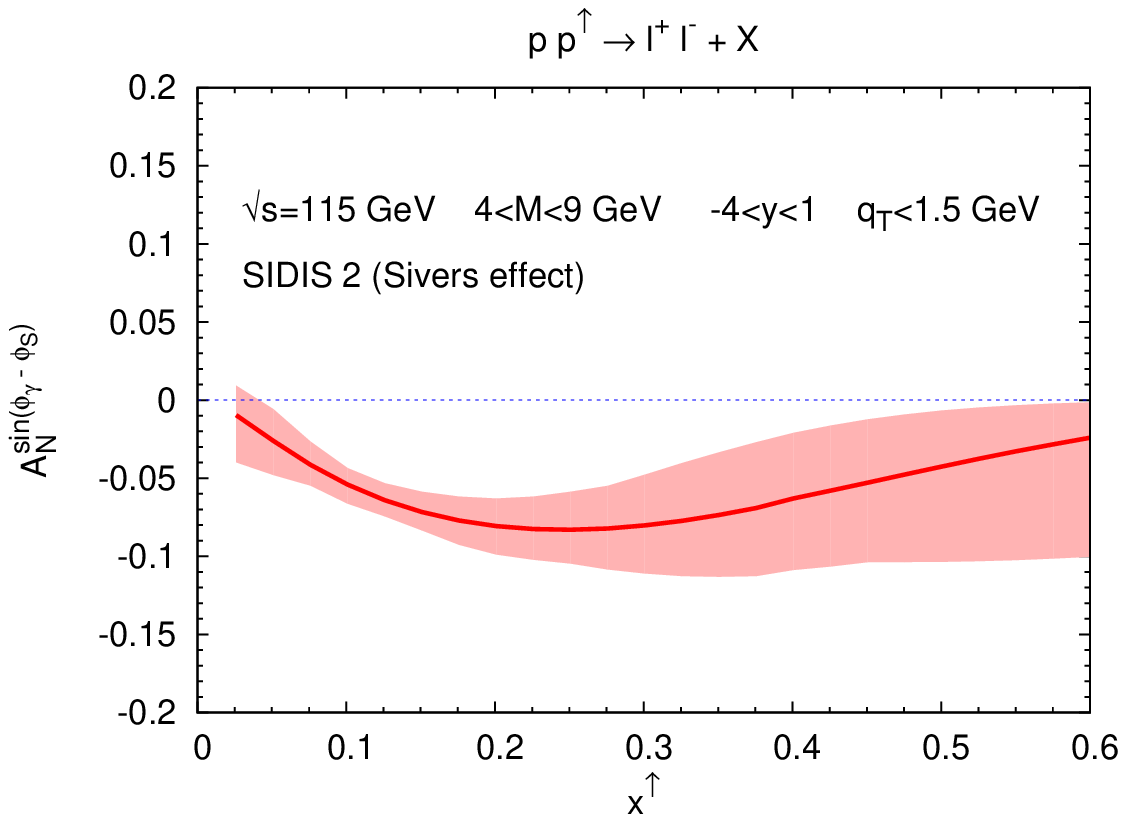}
\caption{Our theoretical estimates for $A_N^{(\phi_\gamma - \phi_S)}$ in
D-Y processes as expected at AFTER@LHC. Our results are presented as function
of $M$ (upper plots), $x_F$ (middle plots) and $x$ of the quark inside the
polarised proton, $x^\uparrow$ (lower plots). The other kinematical variables
are either fixed or integrated, as indicated in each figure.
They are computed according to Ref~\cite{Anselmino:2009st} and
Eq.~(\ref{ab-ba}), adopting the Sivers functions
of Ref.~\cite{Anselmino:2005ea} (SIDIS 1, left panels) and of
Ref.~\cite{Anselmino:2008sga} (SIDIS 2, right panels), {\it reversed in sign}.
The overall statistical uncertainty band, also shown, is obtained following
the procedure described in Appendix A of Ref.~\cite{Anselmino:2008sga}.}
\label{fig10}
\end{figure}

\vspace*{-.28cm}

\acknowledgments
M.A.~and S.M.~acknowledge support from the ``Progetto di Ricerca di Ateneo/CSP" (codice TO-Call3-2012-0103). 
U.D.~is grateful to the Department of Theoretical Physics II of the Universidad Complutense of Madrid for the kind hospitality
extended to him during the completion of this work.\\


\end{document}